\documentclass[aps,pr,onecolumn,superscriptaddress,preprintnumbers,amsmath,amssymb]{revtex4}
\usepackage{graphicx} 
\usepackage{dcolumn}  

\usepackage{epsfig}
\usepackage{epsbox}
\usepackage[T1]{fontenc}
\usepackage{wrapfloat}
\usepackage{subfigure}
\usepackage{here}
\usepackage{array}
\usepackage{amsmath, amssymb, hhline, multirow, hangcaption, subfigure}
\usepackage{graphics}
\usepackage{graphicx}
\usepackage{lscape}
\usepackage{fancyhdr}
\usepackage{longtable}
\usepackage{supertabular}
\usepackage{refmerge}

\def \vec #1{\mbox{{\boldmath $#1$}}}

\def \lr #1{\left( #1 \right)}
\def \suplr #1{^{\left( #1 \right)}}

\def \Diff {{\mathit \Delta}}

\def \abs #1{\left| #1 \right|}

\def \F {{\cal F}}

\def \GeV {{\rm GeV}}
\def \MeV {{\rm MeV}}

\def \eV {{\rm eV}}

\begin{document}



\preprint{\vbox{ 
\hbox{   }
\hbox{  }
}}

\title{ 
High statistics measurement of the cross sections of
$\gamma\gamma\to\pi^+\pi^-$ production}

\begin{abstract}
We report on a high statistics measurement of the total and differential
cross sections of the process $\gamma\gamma\to\pi^+\pi^-$
in the $\pi^+\pi^-$ invariant mass range $0.8~\GeV/c^2 < W < 1.5~\GeV/c^2$
with 85.9~fb$^{-1}$ of data collected at $\sqrt{s}=10.58$~GeV
and 10.52~GeV with the Belle detector.
A clear signal of the $f_0\lr{980}$ resonance is 
observed in addition to the $f_2\lr{1270}$ resonance.
An improved 90\% confidence level upper limit
${\cal B} (\eta ' (958) \to \pi^+ \pi^- ) < 2.9 \times 10^{-3}$ 
is obtained for $P$- and $CP$-violating
decay of the $\eta ' (958)$ meson using the most conservative assumption
about the interference with the background.
\\

\noindent
KEYWORDS: two-photon production, $f_0(980)$, $f_2(1270)$,  $\eta '(958)$,  
$\gamma \gamma \rightarrow \pi^+ \pi^-$ cross section

\end{abstract}


\affiliation{Budker Institute of Nuclear Physics, Novosibirsk}
\affiliation{University of Cincinnati, Cincinnati, Ohio 45221}
\affiliation{Department of Physics, Fu Jen Catholic University, Taipei}
\affiliation{The Graduate University for Advanced Studies, Hayama}
\affiliation{Gyeongsang National University, Chinju}
\affiliation{Hanyang University, Seoul}
\affiliation{University of Hawaii, Honolulu, Hawaii 96822}
\affiliation{High Energy Accelerator Research Organization (KEK), Tsukuba}
\affiliation{Hiroshima Institute of Technology, Hiroshima}
\affiliation{Institute of High Energy Physics, Chinese Academy of Sciences, Beijing}
\affiliation{Institute of High Energy Physics, Vienna}
\affiliation{Institute of High Energy Physics, Protvino}
\affiliation{Institute for Theoretical and Experimental Physics, Moscow}
\affiliation{J. Stefan Institute, Ljubljana}
\affiliation{Kanagawa University, Yokohama}
\affiliation{Korea University, Seoul}
\affiliation{Kyungpook National University, Taegu}
\affiliation{Swiss Federal Institute of Technology of Lausanne, EPFL, Lausanne}
\affiliation{University of Ljubljana, Ljubljana}
\affiliation{University of Maribor, Maribor}
\affiliation{University of Melbourne, School of Physics, Victoria 3010}
\affiliation{Nagoya University, Nagoya}
\affiliation{Nara Women's University, Nara}
\affiliation{National Central University, Chung-li}
\affiliation{National United University, Miao Li}
\affiliation{Department of Physics, National Taiwan University, Taipei}
\affiliation{H. Niewodniczanski Institute of Nuclear Physics, Krakow}
\affiliation{Nippon Dental University, Niigata}
\affiliation{Niigata University, Niigata}
\affiliation{University of Nova Gorica, Nova Gorica}
\affiliation{Osaka City University, Osaka}
\affiliation{Osaka University, Osaka}
\affiliation{Panjab University, Chandigarh}
\affiliation{Peking University, Beijing}
\affiliation{RIKEN BNL Research Center, Upton, New York 11973}
\affiliation{University of Science and Technology of China, Hefei}
\affiliation{Seoul National University, Seoul}
\affiliation{Shinshu University, Nagano}
\affiliation{Sungkyunkwan University, Suwon}
\affiliation{University of Sydney, Sydney, New South Wales}
\affiliation{Toho University, Funabashi}
\affiliation{Tohoku Gakuin University, Tagajo}
\affiliation{Tohoku University, Sendai}
\affiliation{Department of Physics, University of Tokyo, Tokyo}
\affiliation{Tokyo Institute of Technology, Tokyo}
\affiliation{Tokyo Metropolitan University, Tokyo}
\affiliation{Tokyo University of Agriculture and Technology, Tokyo}
\affiliation{Virginia Polytechnic Institute and State University, Blacksburg, Virginia 24061}
\affiliation{Yonsei University, Seoul}
  \author{T.~Mori}\affiliation{Nagoya University, Nagoya} 
  \author{S.~Uehara}\affiliation{High Energy Accelerator Research Organization (KEK), Tsukuba} 
  \author{Y.~Watanabe}\affiliation{Kanagawa University, Yokohama} 
  \author{K.~Abe}\affiliation{Tohoku Gakuin University, Tagajo} 
  \author{I.~Adachi}\affiliation{High Energy Accelerator Research Organization (KEK), Tsukuba} 
  \author{H.~Aihara}\affiliation{Department of Physics, University of Tokyo, Tokyo} 
 \author{K.~Arinstein}\affiliation{Budker Institute of Nuclear Physics, Novosibirsk} 
  \author{V.~Aulchenko}\affiliation{Budker Institute of Nuclear Physics, Novosibirsk} 
  \author{T.~Aushev}\affiliation{Swiss Federal Institute of Technology of Lausanne, EPFL, Lausanne}\affiliation{Institute fo
r Theoretical and Experimental Physics, Moscow} 
  \author{A.~M.~Bakich}\affiliation{University of Sydney, Sydney, New South Wales} 
% ...
\collaboration{The Belle Collaboration}

\date{April 5, 2007}

\maketitle

\tighten

{\renewcommand{\thefootnote}{\fnsymbol{footnote}}}
\setcounter{footnote}{0}

\section{Introduction}
\label{sec:intr}
The nature of low mass mesons remains poorly understood
in spite of decades of theoretical and experimental effort~\cite{bib:scalar}.
In particular, low mass scalar mesons (below 1~$\GeV/c^2$) are not yet
 well established experimentally except for
the $f_0\lr{980}$ and $a_0\lr{980}$ mesons, while the
extensively discussed $\sigma~(f_0(600))$ and $\kappa$~($K^*(800)$) mesons
still remain controversial states~\cite{bib:PDG}.
A $B$ factory is well suited for detailed investigations
of low mass mesons through two-photon production, where overwhelming
statistics can be obtained.
Two-photon production of mesons has advantages over
meson production in hadronic processes;
the production rate can be reliably calculated from QED 
with $\Gamma_{\gamma\gamma}$ as the only unknown parameter.
In addition, a meson can be produced alone without additional hadronic 
debris,
and the quantum numbers of the final state are restricted to states of 
charge conjugation $C=+1$ with $J=1$ forbidden
(Landau-Yang's theorem~\cite{bib:Yang}). 

In the past, extensive studies of low mass mesons through 
$\gamma \gamma \rightarrow \pi \pi$ scattering
have been made at $e^+ e^-$ colliders: 
Crystal Ball~\cite{bib:crysball}, Mark II~\cite{bib:mark2},
JADE~\cite{bib:JADE}, TOPAZ~\cite{bib:TOPAZ},
MD-1~\cite{bib:md1}, CELLO~\cite{bib:CELLO}
and VENUS~\cite{bib:VENUS}; see Ref.~\cite{bib:PDG}
for a list of the earlier experiments.
Using data from Mark II, Crystal Ball, and CELLO,
Boglione and Pennington (BP) performed an amplitude analysis
of $\gamma\gamma\rightarrow\pi^+\pi^-$ and $\gamma\gamma\rightarrow\pi^0\pi^0$
cross sections~\cite{bib:amplitude}.
They found two distinct classes of solutions where
one solution has a peak (``peak'' solution) and the other has a wiggle 
(``dip'' solution) in the $f_0\lr{980}$ mass region.
The two solutions give quite different results
for the two-photon width of the $f_0\lr{980}$ and
the size of the S-wave component.
Thus, it is necessary to distinguish them experimentally.

In this paper, we report on a measurement of the cross sections
for the reaction  $\gamma\gamma\to\pi^+\pi^-$
with high statistics that are more than two orders of 
magnitude larger than that of the past experiments.
The analysis is based on data taken with the 
Belle detector at the KEKB asymmetric-energy (3.5~GeV on 8~GeV) 
$e^+e^-$ collider~\cite{bib:kekb}.
The data sample corresponds to a total integrated luminosity of 85.9~fb$^{-1}$,
accumulated on the $\Upsilon(4S)$ resonance $(\sqrt{s} = 10.58~{\rm GeV})$
and 60~MeV below the resonance (8.6~fb$^{-1}$ of the total).
Since the cross section difference between the two energies is only about
0.3\%, we combine both samples~\cite{bib:lume}.
We observe the two-photon process $e^+e^-\rightarrow e^+e^- \pi^+\pi^-$
in the ``zero-tag'' mode,
where neither the final-state electron nor positron is detected,
and the $\pi^+\pi^-$ system has small transverse momentum. 
We restrict the virtuality of the incident photons to 
be small by imposing a strict requirement on the
transverse-momentum balance of the final-state 
hadronic system with respect to the beam axis.
Some of the results reported here are the subject of
a separate paper focusing on the
properties of the $f_0(980)$ meson~\cite{bib:tmori}.

This paper is organized as follows. 
A brief description of the detector is given in section~\ref{sec:bell}.
The selection criteria are listed in section~\ref{sec:sele}.
There is a well known difficulty in discriminating 
$\mu^{\pm}$ from $\pi^{\pm}$ in the low momentum region 
($\lesssim 0.8~\GeV/c$);
Section~\ref{sec:pid} presents the method of particle identification, 
in particular the method of $\mu / \pi$ separation that we use.
Evaluation of the detection and trigger efficiencies is described in 
section~\ref{sec:eff}.
The total and differential cross sections are given 
while their systematic errors are estimated in section~\ref{sec:cros}.
In section~\ref{sec:fit},
the resulting spectrum is fitted to obtain the resonance parameters
of the $f_0(980)$ meson and to check consistency in the $f_2(1270)$ 
region.
Section~\ref{sec:sum} summarizes the results.
Appendix~\ref{sec:bgds} gives a detailed description of the background 
subtraction.
Values of the total cross sections are given in Appendix~\ref{sec:valu}.

\section{The Belle Detector}
\label{sec:bell}
The Belle detector is a large-solid-angle magnetic spectrometer having good
momentum resolution and particle identification capability
in the energy region of interest~\cite{bib:belle}.
Here we briefly describe the Belle detector components.
Charged track coordinates near the collision point are measured by a
3-layer silicon vertex detector (SVD) that surrounds a 2~cm radius 
beryllium beam pipe.
Track trajectories are reconstructed in a 50-layer central drift chamber (CDC),
and momentum measurements are made together with the SVD.
An array of 1188 silica-aerogel Cherenkov counters (ACC) provides
separation between kaons and pions for momenta above 1.2~GeV/$c$.  
The time-of-flight counter (TOF) system consists 
of a barrel of 128 plastic scintillation counters and is 
effective for $K/\pi$ separation for tracks with momenta below 1.2~GeV/$c$. 
Low energy kaons and protons are also identified through specific
ionization ($dE/dx$) measurements in the CDC.
Photon detection and energy measurements of photons and electrons
are provided by an electromagnetic calorimeter (ECL).
It is comprised of an array of 8736 CsI(Tl) crystals all pointed toward
the interaction point,
greatly enhances the electron identification capability provided through
a comparison of energy measured in the ECL and momentum in the CDC.
These detector components are located within a superconducting 
solenoid coil that provides a uniform magnetic field of 1.5~T.
An iron flux-return located outside the solenoid coil is instrumented
to detect $K^0_L$ mesons and to identify muons (KLM).
The $z$ axis of the detector is defined to be opposite to the direction
of the positron beam.
These detector components cover a polar angular range between $17^{\circ}$
and $150^{\circ}$.

\section{Event Selection}
\label{sec:sele}
Signal candidates are primarily triggered by a two-track trigger 
that requires two CDC tracks with associated TOF hits and ECL clusters 
with an opening angle greater than $135^{\circ}$.
Exclusive $e^+e^-\to e^+e^-\pi^+\pi^-$ events are selected 
by requiring two oppositely charged tracks coming from the interaction 
region; each track is required to satisfy $dr<0.1$~cm and $|dz|<2$~cm, 
where $dr$ ($dz$) is $r$ ($z$) component of the closest approach to the 
nominal collision point.
Here, $r$ is the transverse distance from the $z$ axis.
The difference of the $dz$'s of the two tracks must satisfy the requirement
$|dz_+ - dz_-| \leq 1~{\rm cm}$.
The event must contain one and only one positively charged track
that satisfies $p_t > 0.3~\GeV/c$ and $-0.47 < \cos\theta < 0.82$,
where $p_t$ and $\theta$ are the transverse component of momentum 
and the angle with respect to the $z$-axis.
The scalar sum of the track momenta in each event is required to be less
than $6~\GeV/c$,
and the sum of the ECL energies of the event must be less than
$6~{\rm GeV}$.
Events should not include an extra track with
$p_t > 0.1~\GeV/c$.
The cosine of the opening angle of the tracks must be greater than $-0.997$
to reject cosmic-ray events.
The sum of transverse momentum vectors of the two tracks
$\lr{\sum \vec{p}_t^*}$ should satisfy
$\mid \sum \vec{p}_t^* \mid < 0.1~\GeV/c$; this requirement
separates exclusive two-track events from quasi-real two-photon collisions.

\section{Particle Identification and $\mu/\pi$ separation}
\label{sec:pid}
\begin{figure}[H]
    \centering
   \epsfig{file=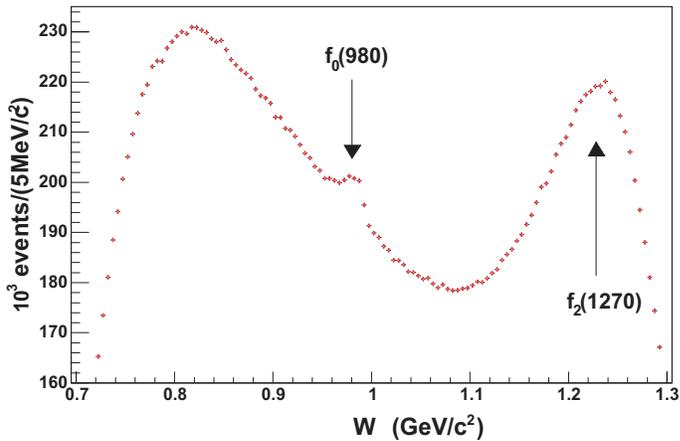,width=90mm}
   \caption{The invariant mass distribution of unseparated 
($\mu^+\mu^- + \pi^+ \pi^-$) events. 
   A clear signal for the $f_0\lr{980}$ can be seen.
   The large peak around $W=1.2~\GeV/c^2$ corresponds to the $f_2\lr{1270}$
   resonance. Note the suppressed zero on the vertical scale.}
    \centering
   \label{fig:fig1}
  \end{figure}

Electrons and positrons are clearly distinguished from hadrons using
the ratio $E/p$, where $E$ is the energy measured in 
the ECL, and $p$ is the momentum from the CDC.
Kaon (proton) candidates are identified using normalized kaon 
(proton) and pion 
likelihood functions ($L_K$ ($L_p$) and $L_{\pi}$, respectively) 
obtained from the particle identification system (combining the
information of the CDC, TOF, ACC and ECL)
with the criterion $L_K/(L_K+L_{\pi})>0.25$ ($L_p/(L_p+L_{\pi})>0.5$), 
which gives a typical identification efficiency of 90\% with a pion 
misidentification probability of 3\%.
All charged tracks that are not identified as electrons, kaons or protons are 
treated as pions.
We require both tracks to be pions.
The resulting invariant mass ($W$) distribution is shown in 
Fig.~\ref{fig:fig1}.
The $W$ bin size is chosen to be 5~$\MeV/c^2$, while the mass resolution
is about 2~$\MeV/c^2$ according to GEANT-3~\cite{bib:geant} based Monte Carlo
(MC) simulation.
A clear signal corresponding to the $f_0(980)$ meson is seen along with the 
well known $f_2(1270)$ resonance.

\begin{figure}[H]
 \subfigure[For $\mu$]
          {\epsfig{file=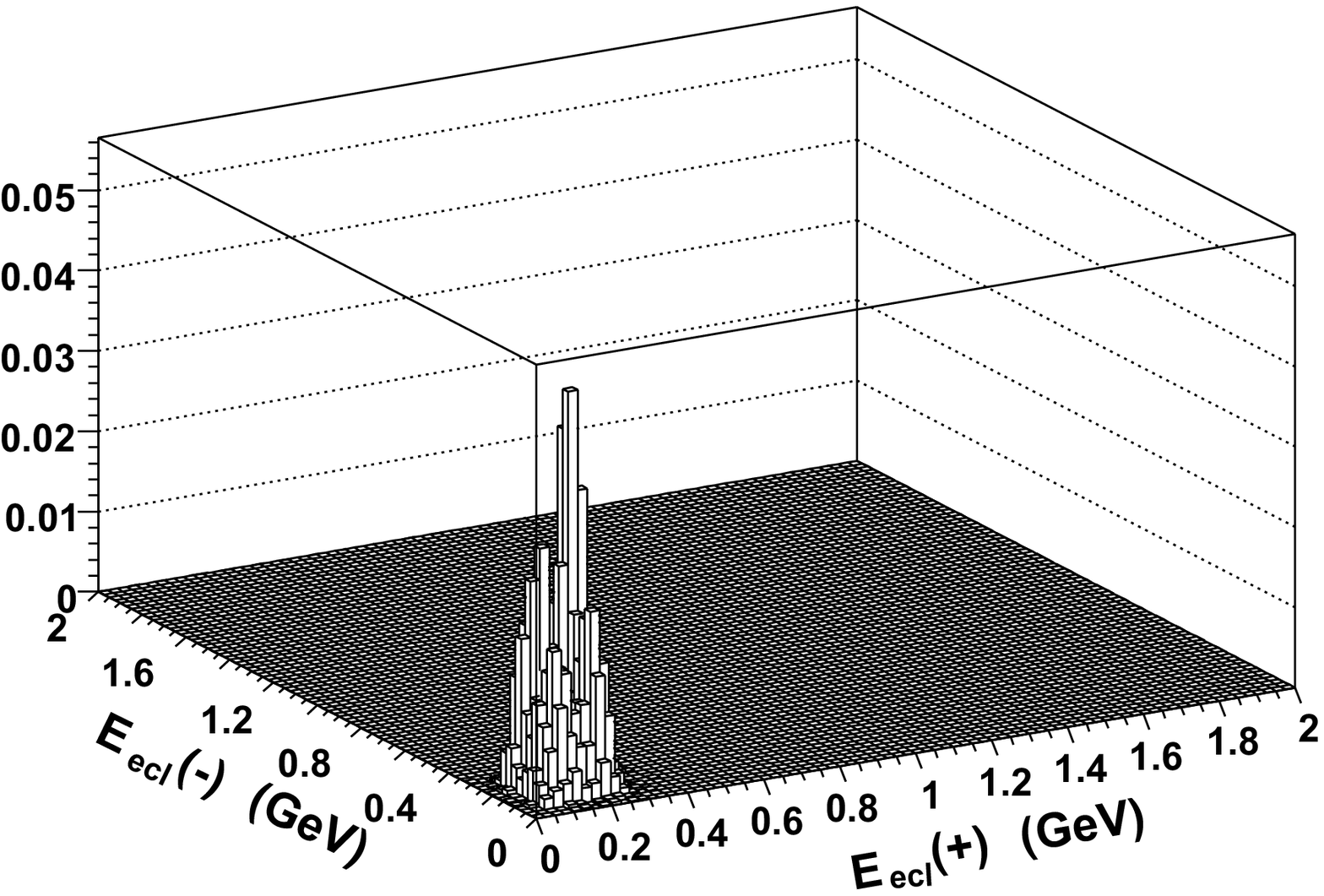,width=70mm}
            \label{fig:fig2a}}
 \subfigure[For $\pi$]
          {\epsfig{file=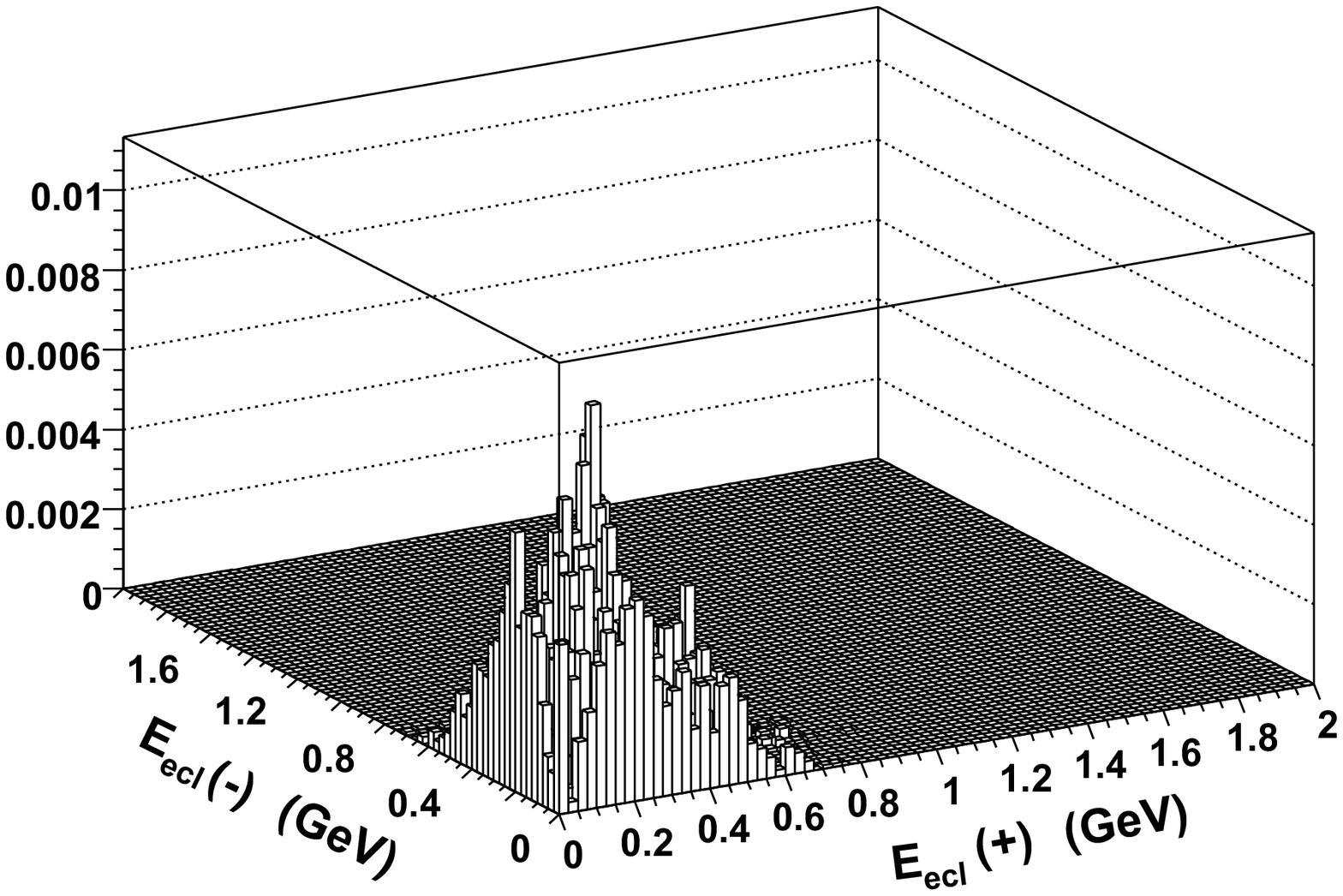,width=70mm}
            \label{fig:fig2b}}
 \caption{Typical distributions of the energy deposit ($E_+$ vs. $E_-$) 
in the ECL. Events are produced using MC simulation in a range
$1.0~\GeV/c^2 < W < 1.02~\GeV/c^2$ and $|\cos \theta^*| < 0.1$.}
\end{figure}

In this measurement, the KLM detector cannot be used for muon identification,
since it is insensitive in the region of interest where the 
transverse momenta of tracks are below $0.8~\GeV/c$.
Therefore, we have developed a method
for separating $\pi^+\pi^-$ and $\mu^+\mu^-$ events statistically using 
ECL information; muons deposit energy corresponding to the 
ionization loss for minimum ionizing particles,
while pions give a wider energy distribution since they may interact 
hadronically in the ECL, which corresponds to approximately one interaction
length of material.
Typical two-dimensional distributions ($E_+$ vs. $E_-$) 
of the energy deposit $E_{\pm}$ in the ECL for $\mu^+ \mu^-$ and 
$\pi^+ \pi^-$ pairs produced by MC are shown in Figs.~\ref{fig:fig2a} 
and \ref{fig:fig2b}.

Probability density functions (PDFs) for the distributions of energy deposits
from $\pi^+ \pi^-$ ($\mu^+\mu^-$) pairs $P_{\pi^+\pi^-}\suplr{i} \lr{E_+,E_-}$
($P_{\mu^+\mu^-}\suplr{i} \lr{E_+,E_-}$)
are obtained with MC simulation.
Here $i$ represents the $i$-th bin of $(W, \abs{\cos\theta^*})$ 
in 20~$\MeV/c^2$ and 0.1 steps,
where $W$ is the invariant mass of the $\pi^+ \pi^-$ (or $\mu^+ \mu^-$)
pair in each event (the pion mass is assumed in the calculation), and
$\theta^*$ is the polar angle of the produced $\pi^\pm$ meson
(or $\mu^\pm$ lepton) in the center-of-mass system of two initial photons. 
Note that using this method the effect of muons from pion decays is taken 
into account by the pion PDFs.
We obtain $r\suplr{i}$, the fraction of $\mu^+ \mu^-$ in the $i$-th bin 
through the equation:
\begin{eqnarray}
N_{\rm data}^{\suplr{i}}\lr{E_+,E_-}
&=& N_{\rm tot}\suplr{i}  \left( 
 r\suplr{i} P_{\mu^+ \mu^-}\suplr{i} \lr{E_+,E_-} +
(1-r\suplr{i})P_{\pi^+ \pi^-}\suplr{i} \lr{E_+,E_-} \right) \; ,
\label{eqn:mupi}
\end{eqnarray}
where $N_{\rm data}^{\suplr{i}}\lr{E_+,E_-}$ is the distribution of data
and $N_{\rm tot}\suplr{i}$ is the total number of events in that bin.
The values of ratios $r\suplr{i}$ obtained must be corrected since the
MC cannot simulate hadronic interactions accurately enough.
By introducing mis-ID probabilities, 
$P_{\pi \pi \rightarrow \mu \mu}$ and $P_{\mu \mu \rightarrow \pi \pi}$, 
the $r$ value for each bin (the bin number $i$ is omitted) 
can be written as:
\begin{equation}
r = \frac{N_{\mu \mu} + N_{\pi \pi} P_{\pi \pi \rightarrow \mu \mu} 
- N_{\mu \mu} P_{\mu \mu \rightarrow \pi \pi}}{N_{\mu \mu} + N_{\pi \pi}} \; ,
\label{eqn:rmupi}
\end{equation}
where $N_{\pi \pi}$ ($N_{\mu \mu}$) is the number of true $\pi^+ \pi^-$
($\mu^+ \mu^-$) pairs in that bin.
We assume that $P_{\pi \pi \rightarrow \mu \mu}$ and 
$P_{\mu \mu \rightarrow \pi \pi}$ are independent of $W$.
Applying the $\mu / \pi$ separation method mentioned above to a sample 
of data events positively identified as muons
by the KLM information in the higher energy region, 
we find that $P_{\mu \mu \rightarrow \pi \pi}$ is statistically
consistent with zero.
The values of $P_{\pi \pi \rightarrow \mu \mu}$ in each $|\cos \theta^*|$ bin 
are determined such that the ratio of the data and MC for 
$\mu^+ \mu^-$ pairs, which is ideally one, gives a straight line in the $W$ 
spectrum. 
The values of $P_{\pi \pi \rightarrow \mu \mu}$ vary between 0.08 to
0.13 in $|\cos \theta^*|$ bins.
Because they are determined for each bin of $|\cos \theta^*|$,
the bin-by-bin variation of systematic errors is rather large in the angular
distribution. 
After subtracting $\mu^+ \mu^-$ events, a total of $6.4 \times 10^6$
events remains in the region of $0.8~\GeV/c^2 < W < 1.5~\GeV/c^2$ and
$|\cos \theta^*| < 0.6$.

\section{Detection and Trigger efficiency}
\label{sec:eff}
The detection (trigger) efficiencies, $\epsilon_{\rm det}$ 
($\epsilon_{\rm trg}$) are estimated from a MC simulation.
Events of the process $\gamma\gamma\to\pi^+\pi^-$ 
are generated using TREPS~\cite{bib:treps}.
The detection efficiency is calculated from the MC simulation
as the ratio of the number of detected and generated events in each bin of 
$W$ (with the bin width, 5~$\MeV/c^2$) and $|\cos\theta^*|$ (0.05).
The MC statistics are high enough and do not contribute to systematic
errors.
\begin{figure}[H]
\centering
 \epsfig{file=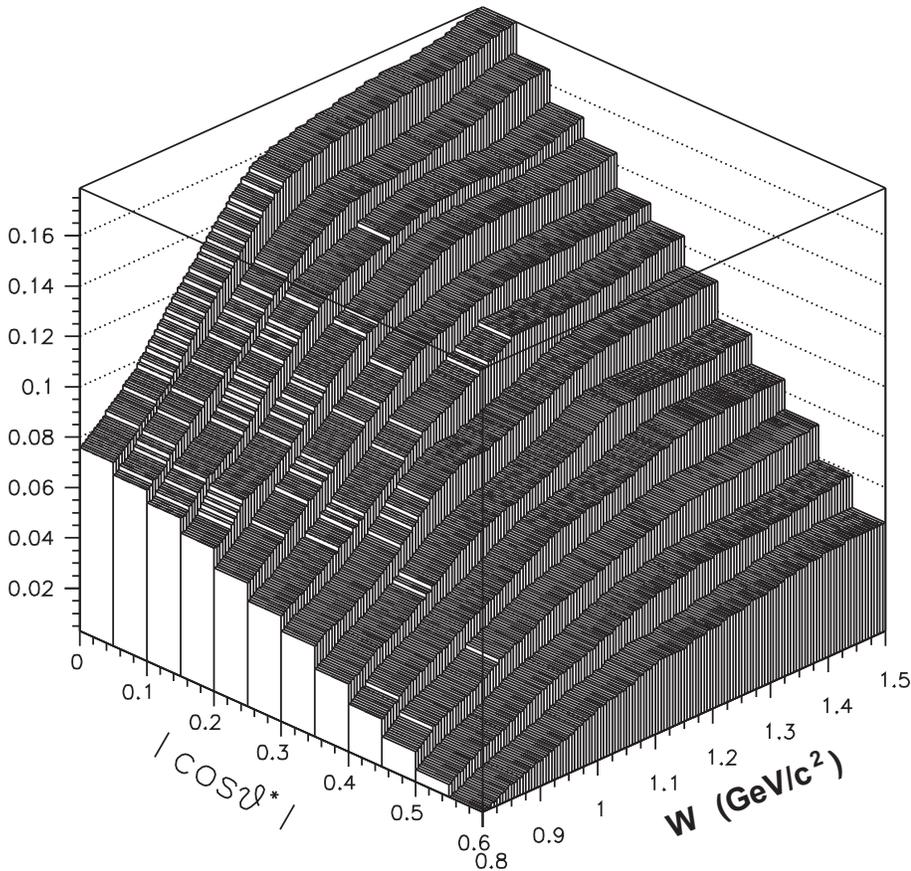,width=120mm}
\centering
 \caption{The combined detection and trigger efficiencies
as a function of $W$ and $|\cos \theta^*|$.}
\label{fig:fig3}
\end{figure}
The trigger efficiencies 
are estimated 
with the same binning using the trigger simulator.
Since the trigger simulator does not simulate triggers very accurately, 
particularly in the low energy region, the efficiency values have to be 
corrected. 
We calculate the correction factors by comparing the number of
$e^+e^-\to e^+e^-e^+e^-$  events in data and MC that are triggered by 
the two-track trigger.
The resulting factors steeply rise from 0.5 at $W=0.8~\GeV/c^2$ to 0.8 
at $W=1~\GeV/c^2$ and then increase gradually for higher $W$.
The combined detection and trigger efficiencies are shown in
Fig.~\ref{fig:fig3}.
The muon-background subtraction and all the correction factors are applied 
using smooth functions obtained by parameterizing the results of bin-by-bin 
analyses.

\section{Cross Sections}
\label{sec:cros}
In this section, we derive differential and total cross sections
and evaluate systematic errors.

\subsection{Differential Cross Sections}
Differential cross sections for $\gamma\gamma\to\pi^+\pi^-$ 
are evaluated by using the following relation:
\begin{equation}
\frac{\Diff \sigma_{\gamma\gamma\to\pi^+\pi^-}}{\Diff |\cos \theta^*|}
  = \frac{\Diff N_{e^+e^-\to e^+e^-\pi^+\pi^-}}
         {\epsilon_{\rm trg}\cdot \epsilon_{\rm det} \cdot 
\Diff W \cdot \Diff |\cos \theta^*| \cdot 
\frac{d{\cal L}}{dW}\cdot \int L dt} \; , 
\label{eqn:dsdcos}
\end{equation}
where $\Diff N_{e^+e^-\to e^+e^-\pi^+\pi^-}$ is the number of events
in a $W$-$|\cos \theta^*|$ bin, $\frac{d{\cal L}}{dW}$ is the two-photon 
luminosity function~\cite{bib:lum_func}
and $\int L dt = 85.9~{\rm fb}^{-1}$ is the integrated luminosity.
Here the $W$ and $|\cos \theta^*|$ bin sizes are also chosen to be 
$5~\MeV/c^2$ and 0.05, respectively.
Background from $\eta ' (958) \to \rho^0 \gamma \to \pi^+ \pi^- \gamma$
is subtracted, a detailed account of which is given in 
Appendix~\ref{sec:bgds}.
The contribution of the background to the cross section is about 5\% at
0.8~$\GeV/c^2$ and dies away quickly to zero above 0.9~$\GeV/c^2$.
Other backgrounds are negligible.

\begin{figure}[H]
 \centering
   {\epsfig{file=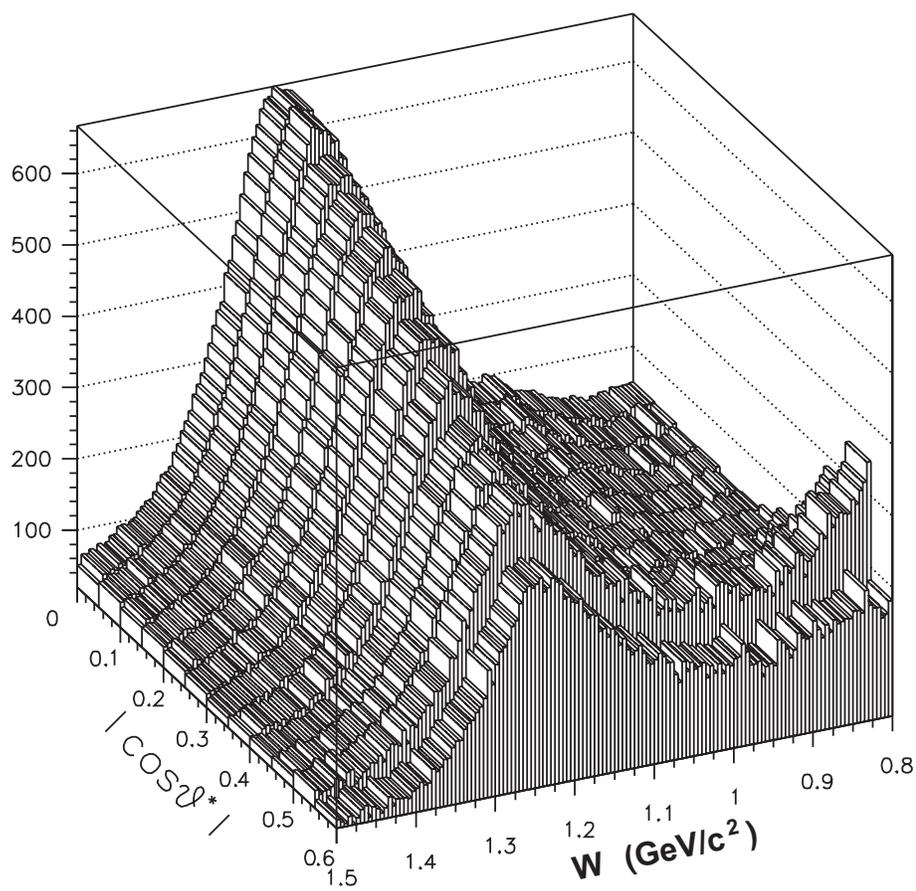,width=120mm}}
 \caption{Differential cross sections 
($d \sigma / d |\cos \theta^*|$ (nb)).
(The $W$-axis is reversed compared to that of Fig.~\ref{fig:fig3}
so as to allow a clearer view of the region in $W$ above the $f_2(1270)$ 
resonance.)}
\label{fig:fig4}
\end{figure}

\begin{figure}[H]
            {\epsfig{file=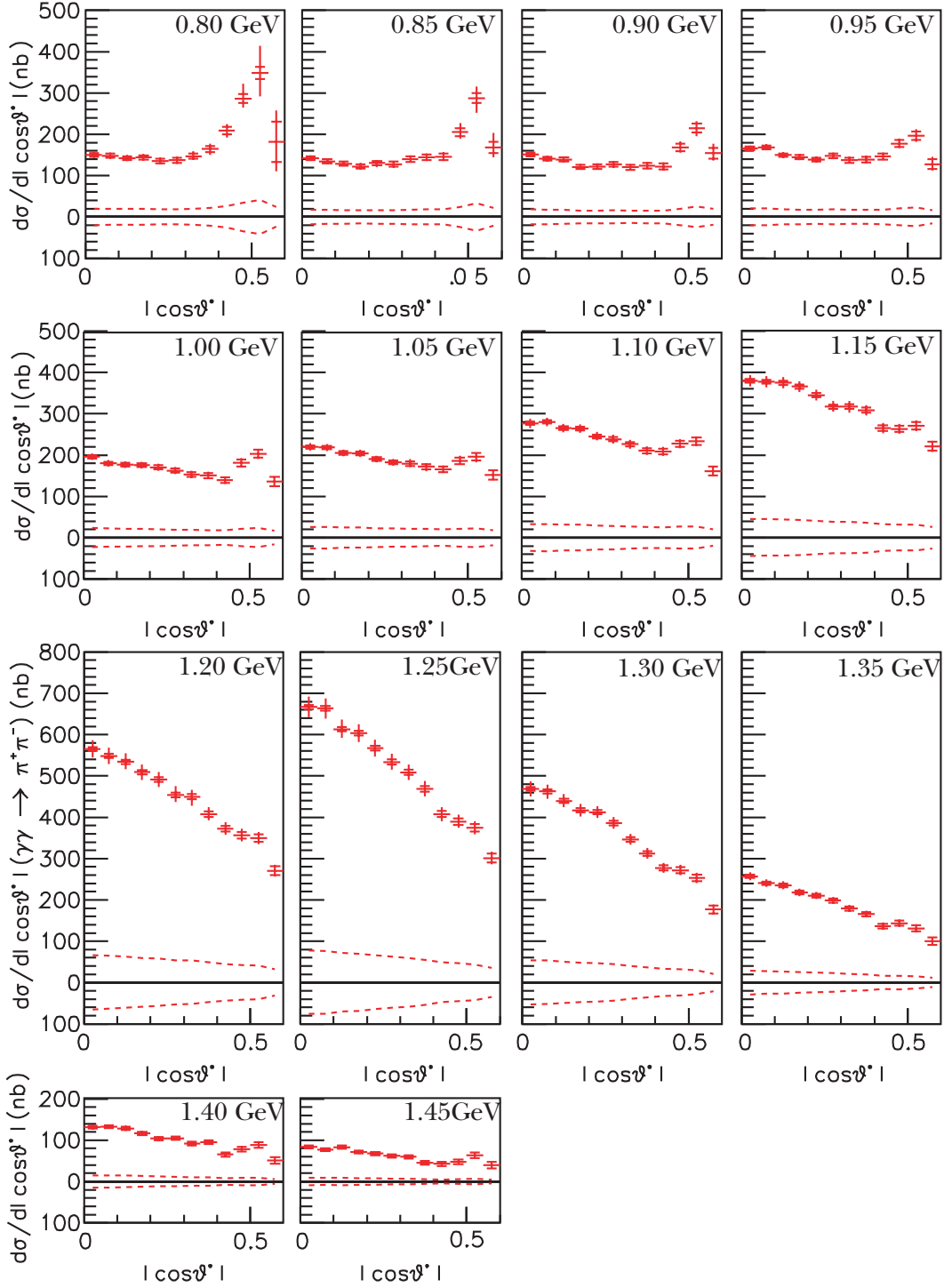,width=140mm}}
 \caption{Representative differential cross sections 
($d \sigma / d |\cos \theta^*|$ (nb)).
In the figures, 0.80~GeV means a bin of
$0.800~\GeV/c^2 < W < 0.805~\GeV/c^2$, etc., 
and the dashed lines indicate the upper and lower overall systematic errors.
The two short horizontal bars indicate the statistical errors while the
vertical ones include point-by-point systematic errors.}
\label{fig:fig5}
\end{figure}

Differential cross sections $d \sigma / d |\cos \theta^*|$
are obtained using Eq.~(\ref{eqn:dsdcos}) for $|\cos \theta^*|$ 
from 0 to 0.6
and for $W$ from 0.8~$\GeV/c^2$ up to 1.5~$\GeV/c^2$.
The resulting differential cross sections
are shown in Fig.~\ref{fig:fig4}.
In order to present the cross sections more quantitatively,
some representative ones are also plotted in Fig.~\ref{fig:fig5}.
Both statistical and point-by-point errors are shown.
The latter come from the $\mu/\pi$ separation method and trigger efficiency
corrections we employ as explained in Sections~\ref{sec:pid} and 
~\ref{sec:eff}.
A point-by-point systematic error is taken to be one half of the difference 
between the corrections in neighboring bins.
The result shows some apparent systematic structure in the region 
$0.45 < |\cos \theta^*| < 0.6$, particularly for $W < 1.1~\GeV/c^2$.
As shown below (see Eq.~(\ref{eqn:diff})), the differential cross sections 
in this $W$ region can be described by a second order polynomial in 
$|\cos \theta^*|^2$.
Thus, such structures are not considered to be real; 
either the last points are too low or earlier points are too high.
However, we have not identified the cause of the measurement bias.

\begin{figure}
 \centering
   {\epsfig{file=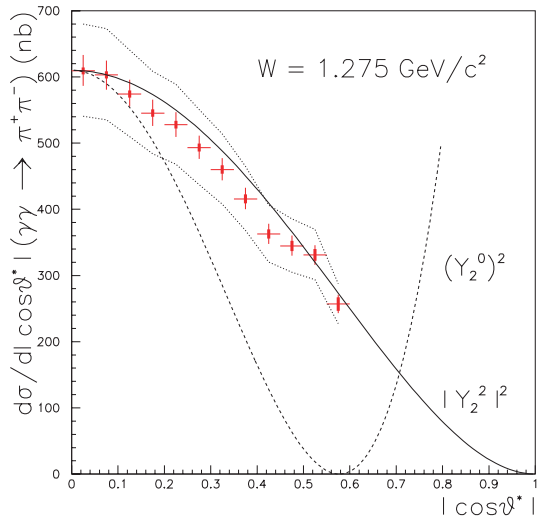,width=70mm}}
 \centering
 \caption{The differential cross section of $\gamma \gamma \to \pi^+\pi^-$ 
at the $f_2(1270)$ mass (1.275~GeV/$c^2$).
Thick vertical bars show statistical errors and thin ones
include point-by-point errors.
The dotted lines indicate the overall systematic errors.
The solid line shows the angular dependence of $Y^2_2$
and the dashed one shows that of $Y^0_2$ (both normalized at 
$\cos \theta^* =0$). }
\label{fig:fig6}
\end{figure}

In this $W$ region, $J > 2$ partial waves (the next one is $J=4$) may be 
neglected so that only S and D waves are to be considered.
The differential cross section can be expressed as:
\begin{equation}
\frac{d \sigma}{d \Omega} (\gamma \gamma \to \pi^+ \pi^-)
 = \left| S \: Y^0_0 + D_0 \: Y^0_2 \right|^2 
+ \left| D_2 \: Y^2_2 \right|^2 \; ,
\label{eqn:diff}
\end{equation}
where $D_0$ ($D_2$) denotes the helicity 0 (2) component of the D wave
and $Y^m_J$ are the spherical harmonics:
\begin{equation}
Y^0_0 = \sqrt{\frac{1}{4 \pi}}, \;
Y^0_2 = \sqrt{\frac{5}{16 \pi}}(3 \cos^2 \theta^* - 1), \;
\left| Y^2_2 \right| = \sqrt{\frac{15}{32 \pi}} \sin^2 \theta^* \; .
\label{eqn:sphe}
\end{equation}
Since $|Y^2_2|$ is not independent of $Y^0_2$ and $Y^0_0$ 
(i.e. $|Y^2_2| = (\sqrt{5} Y^0_0 - Y^0_2)/\sqrt{6}$), partial waves
cannot be separated from the differential cross 
sections alone; additional inputs or assumptions are needed.
The general trend of the angular distribution as a function of $W$ is 
as follows.
The angular distribution below $1~\GeV/c^2$ is rather flat for 
$|\cos \theta^*| \le 0.4$, indicating that
the S wave fraction is significant (Fig.~\ref{fig:fig5}).
In the region above $1~\GeV/c^2$, the angular dependence becomes steeper
as $W$ increases and is the steepest around the $f_2(1270)$ mass.
Such behavior is typical of D wave dominance.
Theoretically, the helicity=2 wave ($D_2$) is expected to be 
dominant~\cite{bib:hel2}.
This is supported by Fig.~\ref{fig:fig6}, where
the angular dependence of $|Y^2_2|^2$ and $(Y^0_2)^2$ is plotted
at the $f_2(1270)$ mass.

\subsection{Total Cross Section}
\begin{figure}
 \centering
 \epsfig{file=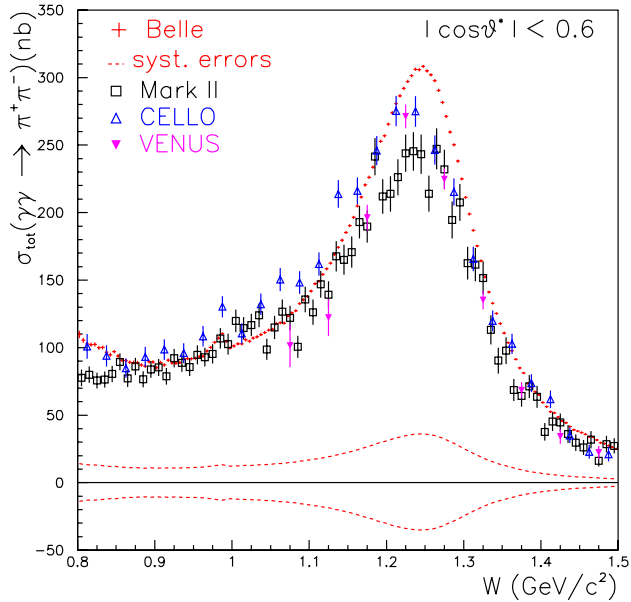,width=90mm}
 \centering
 \caption{The total cross section of $\gamma \gamma \to \pi^+\pi^-$
between 0.8 and 1.5~GeV/$c^2$ for $|\cos \theta^*|<0.6$. 
The Belle data are represented by crosses with statistical error bars, 
the Mark II data are squares,
the CELLO data are the open triangles and the VENUS data are the filled
triangles. 
Dashed lines indicate upper and lower systematic uncertainties for 
the Belle data. 
Numerical values are listed in Appendix~\ref{sec:valu}.
We do not show systematic errors for the other experiments; they are of
similar size or larger.}
\label{fig:fig7}
\end{figure}

The total cross section is then obtained by integrating the differential cross 
sections over $|\cos \theta^*|$ up to 0.6 and
is shown in Fig.~\ref{fig:fig7}
together with the results of some past experiments.
A clear peak corresponding to the $f_0(980)$ meson is visible, indicating
that the peak solution of the BP analysis is preferred. 
Systematic errors for the total cross section are summarized in 
Table~\ref{tab:syserr_sigma}.
They are dominated by the uncertainty in the $\mu/\pi$ separation and that
of the trigger efficiency.
Systematic errors arising from the $\mu/\pi$ separation are estimated by 
changing the value of $P_{\pi \pi \rightarrow \mu \mu}$ in the allowable range 
in each angular bin. 
Since  $\mu^+ \mu^-$ events are well identified by the KLM for $W > 1.6$~GeV,
the allowable range is determined in this region.
These well identified $\mu^+ \mu^-$ events are also used in estimating 
systematic errors of the trigger efficiency.
Comparing data and MC for $\mu^+ \mu^-$ events in the region $W > 1.6$~GeV and 
extrapolating linearly downward, the systematic errors are found to be 4\% 
at $W=1.5$~GeV and 10\% at $W=0.8$~GeV.
The total systematic error is obtained by summing the systematic errors
in quadrature and is also shown in Fig.~\ref{fig:fig7}.
Our results are in good agreement with past experiments
except for the $f_2\lr{1270}$ mass peak region, where our data points are
about 10 to 15\% larger, but still within the systematic errors.
\begin{center}
 \begin{table}[h]
  \caption{Summary of systematic errors
           for the $\gamma\gamma\to\pi^+\pi^-$ cross section.
           A range is shown when the uncertainty has $W$ dependence.}
  \label{tab:syserr_sigma}
  \begin{tabular}{lc}
   \hline\hline 
   Parameter & Syst. error (\%) \\
   \hline
   Tracking efficiency   & 2.4     \\
   Trigger efficiency    & 4 -- 10  \\
   $K/\pi$-separation    & 0 -- 1       \\
   $\mu/\pi$-separation  & 5 -- 7   \\
   Luminosity function   & 5     \\
   Integrated luminosity & 1.4       \\
   \hline
   Total                 & 11.1 -- 12.3 \\
   \hline\hline 
  \end{tabular}
 \end{table}
\end{center}
\vspace{-\baselineskip}

\section{Fits to the Cross Sections}
\label{sec:fit}
The results of the cross section measurements can be used to obtain
the parameters of $f_0(980)$ and $f_2(1270)$ resonances, and to search for
other states decaying into $\pi^+ \pi^-$.  
Some of us plan to perform a full amplitude analysis in the near future
using the present data including the differential cross sections along
with published cross section data of the past.
Thus, we restrict our analysis to a simple level in this paper.
In this section, we summarize the measurement of the parameters of 
the $f_0(980)$ discussed in a separate paper~\cite{bib:tmori}, 
perform a simple fit for the $f_2(1270)$ resonance as a
consistency check, and search for $P$- and $CP$-violating decay of the
$\eta '(958)$ meson into a $\pi^+ \pi^-$ pair.

\subsection{The $f_0\lr{980}$ Resonance}
We have to take into account the effect of the $K\bar{K}$ channel that opens
within the $f_0\lr{980}$ mass region.
The fitting function for the scalar resonance $f_0(980)$ is
parameterized as follows:
\begin{equation}
 \sigma = \abs{ \F^{f_0}e^{i \varphi} +
 \sqrt{\sigma^{\rm BG}_0}}^2
  + \sigma^{\rm BG} - \sigma^{\rm BG}_0 ,
 \label{eqn:sigma}
\end{equation}
where 
$\F^{f_0}$ is the amplitude of the $f_0(980)$
meson~\cite{bib:norm}, which interferes with the helicity-0-background 
amplitude $\sqrt{\sigma^{\rm BG}_0}$ with a relative phase $\varphi$, and 
$\sigma^{\rm BG}$ is the total background cross section.
The amplitude $\F^{f_0}$ can be written as 
\begin{equation}
\F^{f_0} = \frac{\sqrt{4.8 \pi \beta_{\pi}}}{W} \cdot
\frac{g_{f_0 \gamma\gamma}g_{f_0 \pi\pi}}{16\pi}
\cdot\frac{1}{D_{f_0}} ,
\label{eqn:ff0}
\end{equation}
where the factor 4.8 includes the fiducial angular acceptance
$|\cos \theta^*|<0.6$,
$\beta_X = \sqrt{1-\frac{4 {m_X}^2}{W^2}}$ is the velocity of the 
particle $X$ with mass $m_X$ in the two-body final state $X \bar{X}$,
and $g_{f_0XX}$ is related to the partial width 
of the $f_0(980)$ meson via
$\Gamma_{XX} (f_0) = \frac{\beta_X g_{f_0 XX}^2}{16 \pi m_{f_0}}$.
The factor $D_{f_0}$ is given as follows~\cite{bib:denom}:
\begin{eqnarray}
 D_{f_0}(W) &=& m_{f_0}^2 - W^2
              + \Re{\Pi_{\pi}^{f_0}}\lr{m_{f_0}}-\Pi_{\pi}^{f_0}\lr{W}
+ \Re{\Pi_K^{f_0}}\lr{m_{f_0}} - \Pi_K^{f_0}\lr{W} ,
\label{eqn:df0}
\end{eqnarray}
where for $X = \pi$ or $K$,
$\Re{\Pi_X^{f_0}}\lr{m_{f_0}}$ is the real part of
$\Pi_X^{f_0}\lr{m_{f_0}}$, which is given by:
\begin{equation}
 \Pi_X^{f_0}(W) =  \frac{\beta_X {g^2_{f_0XX}}}{16\pi}
       \left[i + \frac{1}{\pi}
			          \ln\frac{1-\beta_X}
				          {1+\beta_X}\right] .
\label{eqn:pif0}
\end{equation}
The phase factor $\beta_K$ is real in the region $W \geq 2m_K$
and becomes imaginary for $W < 2m_K$.
The mass difference between  $K^{\pm}$ and $K^0$ $(\bar{K}^0)$ is 
included by taking $\beta_K = \frac{1}{2} (\beta_{K^{\pm}} + \beta_{K^0})$.

\begin{figure}
 \centering
 \epsfig{file=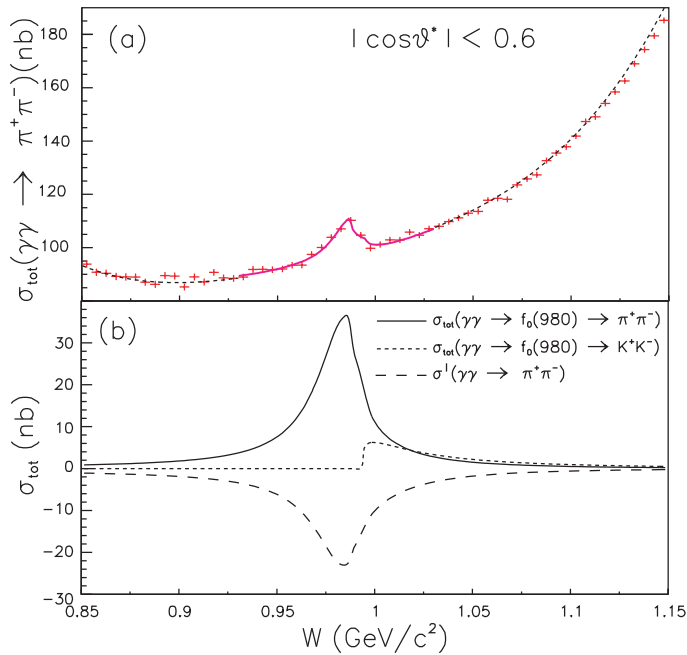,width=90mm}
 \centering
 \caption{Results of the fit: (a) the total cross section (the solid curve)
 (b) contributions of the resonance 
($\sigma(\gamma \gamma \rightarrow f_0(980) \rightarrow \pi^+ \pi^-))$
(solid line) and the interference (dashed).
The cross section of 
 $\sigma(\gamma \gamma \rightarrow f_0(980) \rightarrow K^+ K^-)$
is also shown (dotted)~\cite{bib:tmori}.}
 \label{fig:fig8}
\end{figure}

The results of the fit (shown in Fig.~\ref{fig:fig8} and in 
Table~\ref{tab:f0fit}) are discussed 
in a separate paper~\cite{bib:tmori}.
For completeness we report here the parameters of the $f_0\lr{980}$ 
meson obtained from the fit. 
\begin{eqnarray}
 m_{f_0} &=& 985.6 ~^{+1.2}_{-1.5}\lr{\rm stat}
                   ~^{+1.1}_{-1.6}\lr{\rm syst}~\MeV/c^2
\nonumber \\
 \Gamma_{\pi^+\pi^-}\lr{f_0} &=& 34.2~^{+13.9}_{-11.8}\lr{\rm stat}
                             ~^{+8.8}_{-2.5}\lr{\rm syst}~\MeV \nonumber \\
 \Gamma_{\gamma\gamma}\lr{f_0} &=& 205
                                   ~^{+95}_{-83}\lr{\rm stat}
                                   ~_{-117}^{+147}\lr{\rm syst}~\eV.
\nonumber 
\end{eqnarray}
The two-photon width given by the PDG~\cite{bib:PDG} is
$ \Gamma_{\gamma\gamma}\lr{f_0} = 310~_{-110}^{+80}\lr{\rm stat}~\eV$, and
the value found by BP is $280^{+90}_{-130}$~eV.
Our value of the two-photon width is consistent with them within errors.
\begin{center}
\begin{table}[H]
\caption{Fitted parameters of the $f_0\lr{980}$ region to 
Eq.(\ref{eqn:sigma}).}
\label{tab:f0fit}
\begin{tabular}{llll} \hline \hline
parameter~~~~~~~~~~  & value~~~~~ & \multicolumn{2}{c}{error} \\ 
&& stat~~~~~~~~~ & syst \\ \hline
$m_{f_0}$ ($\MeV/c^2$)  & 985.6 & $^{+1.2}_{-1.5}$ & $^{+1.1}_{-1.6}$ \\
$g_{\pi \pi}$ (GeV)  & 1.33 & $^{+0.27}_{-0.23}$ & $^{+0.16}_{-0.05}$ \\
$\Gamma_{\gamma \gamma} (f_0)$ (eV)  & 205 &$^{+95}_{-83}$
& $^{+147}_{-117}$ \\
$\sigma_0^{\rm BG}$ (nb)  & 3.7 & $^{+1.2}_{-1.5}$ & $^{+4.3}_{-3.9}$ \\
$\varphi$ (rad)  &  1.74 & $\pm 0.09$ & $^{+0.04}_{-0.34}$ \\ \hline
$\chi^2/ndf \; (ndf)$ & 0.90 (15)& &  \\
\hline\hline
\end{tabular}
\end{table}
\end{center}

\subsection{The $f_2\lr{1270}$ Region}
From the past experiments~\cite{bib:crysball, bib:mark2, bib:JADE, bib:TOPAZ,
bib:md1, bib:CELLO, bib:VENUS},
it is well known that the position of the $f_2(1270)$ resonance peak in 
two-photon production is shifted to lower mass because of interference 
with non-resonant background~\cite{bib:f2shift}.
In this paper, we give the result of a simple fit made as a consistency check
in the $f_2(1270)$ region.
The relativistic Breit-Wigner resonance amplitude $A_R(W)$ for a 
spin-$J$ resonance $R$ of mass $m_R$ is given by
\begin{eqnarray}
A_R^J(W) &=& \sqrt{\frac{8 \pi (2J+1) F_J m_R}{W}} 
\times \frac{\sqrt{ \Gamma_{\gamma \gamma}(W) \Gamma_{\pi^+ \pi^-}(W)}}
{m_R^2 - W^2 - i m_R \Gamma_{\rm tot}(W)} \; ,
\label{eqn:arj}
\end{eqnarray}
where $F_J$ is the factor coming from the limited solid angle 
($| \cos \theta^*| <0.6$). 
Hereafter we consider the case $J=2$ (the $f_2(1270)$ meson).
The factor $F_2 = 0.884$ is obtained assuming helicity-2 
dominance~\cite{bib:hel2}; 
the angular dependence is assumed to be $Y^2_2$.
The energy-dependent total width $\Gamma_{\rm tot}(W)$ is given by
\begin{equation}
\Gamma_{\rm tot}(W) = \sum_X \Gamma_{X \bar{X}} (W) \; ,
\label{eqn:gamma}
\end{equation}
where $X$ is $\pi$, $K$, $\gamma$, etc.
The partial width $\Gamma_{X \bar{X}}(W)$ is 
parameterized as~\cite{bib:blat}:
\begin{equation}
\Gamma_{X \bar{X}} (W) = \Gamma_R {\cal B}(R \rightarrow X \bar{X}) 
\left( \frac{q_X(W^2)}{q_X(m_R^2)} \right)^5
\frac{D_2\left( q_X(W^2) r_R \right)}{D_2 \left( q_X(m_R^2) r_R \right)} \;,
\label{eqn:gamx}
\end{equation}
where $\Gamma_R$ is the total width at the resonance mass,
$q_X(W^2) = \sqrt{W^2/4 - m_X^2}$, $D_2(x) = 1/(9 + 3 x^2 +x^4)$,
and $r_R$ is an effective interaction radius that varies from 1~$\GeV^{-1}$ 
to 7~$\GeV^{-1}$ in different hadronic reactions~\cite{bib:grayer}.
For $X = \pi, \; K, \;{\rm and} \; \gamma$,
the branching fractions are $0.848^{+0.025}_{-0.013}$,
$0.046 \pm 0.004$, and $(1.41 \pm 0.13) \times 10^{-5}$, 
respectively~\cite{bib:PDG}.
For the $4 \pi$ and the other decay modes,
$\Gamma_{4 \pi} (W) = \Gamma_R {\cal B}(R \rightarrow 4 \pi)
\frac{W^2}{m_R^2}$ is used instead of Eq.~(\ref{eqn:gamx}).

The fitting function for the $f_2(1270)$ region is taken to be as follows:
\begin{equation}
\sigma = \abs{A_R^{J=2} (W) e^{i \phi_2} + 
b_0 \left( \frac{W}{1~\GeV/c^2} \right)^{-b_1}}^2  + 
c_0 + c_1 W + c_2 W^2  \; , 
\label{eqn:sigma_f2}
\end{equation}
where the contribution other than that of the $f_2(1270)$ resonance
is subdivided into the interfering part (helicity=2) 
and the non-interfering part (helicity=0).
The fit region is chosen to be $\pm \Gamma_{\rm tot}$ around the $f_2$ mass,
i.e. $1.090~\GeV/c^2 < W < 1.461~\GeV/c^2$.
The parameters of the $f_2(1270)$ meson are fixed to the values from the PDG:
the branching fractions as listed above,
$m_R = 1275.4 \pm 1.1~\MeV/c^2$
and $\Gamma_R = 185.2 ^{+3.1}_{-2.5}$~MeV~\cite{bib:PDG},
and the parameter $r_R$ is floated.

\begin{figure}[H]
 \centering
 \epsfig{file=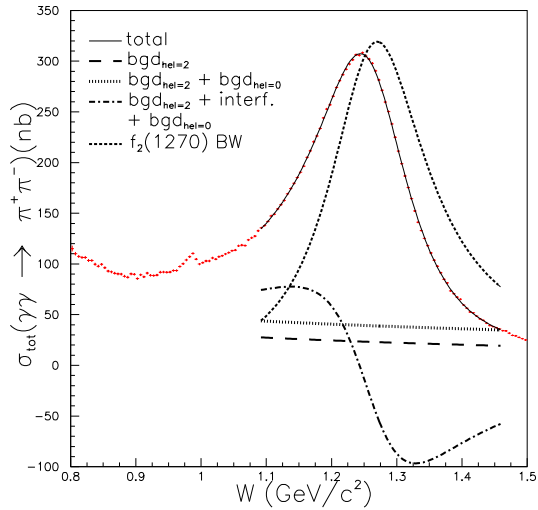,width=70mm}
 \centering
 \caption{Results of the fit of the $f_2(1270)$ region
to Eq.~(\ref{eqn:sigma_f2}). 
The parameters of the $f_2(1270)$ mesons are fixed to the values by the PDG 
(with the helicity=0 background).}
\label{fig:fig9}
\end{figure}

The result of the fit is shown in Fig.~\ref{fig:fig9} and the
obtained parameters are summarized in Table~\ref{tab:f2fit}, where
errors shown are statistical only.
Since a good fit is obtained with $c_2 = 0$, we omit $c_2$. 
A fit without the non-interfering background gives much worse results
as summarized in Table~\ref{tab:f2fit}.
We conclude that the consistency check is satisfactory.
\begin{center}
\begin{table}[H]
\caption{Fitted parameters for the $f_2\lr{1270}$ region
to Eq.~(\ref{eqn:sigma_f2}). 
The parameters of the $f_2(1270)$ meson are fixed to the values by the PDG.
Errors shown are statistical only.}
\label{tab:f2fit}
\begin{tabular}{lcc} \hline \hline
parameter  &  ~~with hel.=0 bgd~~ & ~~without hel.=0 bgd~~\\ \hline
$r_R$ (GeV$^{-1}$)  & $3.62 \pm 0.03$ & $2.84 \pm 0.04$ \\
$b_0$ ($\sqrt{\rm nb}$) & $5.54 \pm 0.02$ & $7.70 \pm 0.05$\\
$b_1$  & $0.61 \pm 0.05$ & $2.00 \pm 0.02$ \\
$\phi_2$ (deg.)  & $28.7 \pm 0.2$ & $22.8 \pm 0.1$ \\
$c_0$ (nb) & $18.2 \pm 3.7$ & 0 (fixed) \\
$c_1$ (nb/$\GeV/c^2$) & $-1.7 \pm 2.8$ & 0 (fixed)\\
$c_2$ (nb/$(\GeV/c^2)^2$) & 0 (fixed) & 0 (fixed)\\ \hline
$\chi^2/ndf \; (ndf)$ & 1.1 (68) & 2.1 (70) \\
\hline\hline
\end{tabular}
\end{table}
\end{center}

\subsection{The $\pi^+\pi^-$ branching fraction of the $\eta ' \lr{958}$ meson}
The $\eta ' \lr{958}$ meson is a pseudoscalar meson and, thus, its coupling
to $\pi \pi$ violates $P$ and $CP$. 
The present upper limit for the $\pi^+\pi^-$ branching fraction 
${\cal B} (\eta ' \rightarrow \pi^+ \pi^-)$ is 2\%~\cite{bib:ritten}.
The high statistics data of Belle allow for a more sensitive search.
The $\eta ' \lr{958}$ meson has a small width of 
$\Gamma_{\eta '} = 0.202 \pm 0.016~\MeV$
and a mass of $m_{\eta '} = 957.78 \pm 0.14~\MeV/c^2$.
Thus its contribution to the $W$ spectrum
can be represented by a Gaussian function:
\begin{equation}
f_{\eta '} (W) dW = \frac{S_{\eta '}}{\sqrt{2 \pi} \sigma_W}
\exp \left( - \frac{(W - m_{\eta '})^2}{2 \sigma_W^2} \right) dW \; ,
\label{eqn:gauss}
\end{equation}
where $S_{\eta '}$ is the parameter to be determined, and
$\sigma_W = 2.0 \pm 0.2~\MeV/c^2$ is the mass resolution determined from MC.
The total cross section in the region $0.92~\GeV/c^2 \le W \le 0.98~\GeV/c^2$
is fitted with a second-order polynomial plus Eq.~(\ref{eqn:gauss}).
The finite bin-size effect is taken into account by integrating the
Gaussian over each bin.
The result is $S_{\eta '} = -27 \pm 16~{\rm nb}\cdot\MeV$.
The systematic error is found to be negligible, which is estimated by 
constraining the $\eta '$ mass and the mass resolution within
one standard deviation and by changing the fitting region.

The parameter $S_{\eta '}$ can be related to the $\pi^+ \pi^-$ 
branching fraction ${\cal B}(\eta ' \to \pi^+ \pi^-)$ as follows.
The cross section formula to be used is the same as Eq.~(\ref{eqn:sigma})
except for replacing the amplitude
${\cal F}^{f_0}$ by Eq.~(\ref{eqn:arj}) with $J=0$:
\begin{equation}
\sqrt{\frac{4.8 \pi m_{\eta '}}{W}}
\frac{ \sqrt{ \Gamma_{\eta '} \Gamma_{\gamma \gamma} 
{\cal B}(\eta ' \to \pi^+ \pi^-)}} 
{M^2_{\eta '} -W^2 -i m_{\eta '} \Gamma_{\eta '}}
\simeq - \frac{ \sqrt{ 4.8 \pi \Gamma_{\eta '}
 \Gamma_{\gamma \gamma}{\cal B}(\eta ' \to \pi^+ \pi^-)} }
{2 m_{\eta '} \left( W - m_{\eta '} + i  \frac{\Gamma_{\eta '}}{2}\right) }
 \; ,
\label{eqn:geta}
\end{equation}
where $\Gamma_{\gamma \gamma} = 4.30 \pm 0.15$~keV~\cite{bib:PDG}
is the two-photon width of the $\eta '$ meson, and
the latter equation is obtained in a narrow width approximation.
Taking into account an interference effect (for the most conservative case)
 and using the relation
$\int_0^{\infty}
d W / ((W - m_{\eta '})^2 + \Gamma_{\eta '}^2/4 )
\simeq 2 \pi / \Gamma_{\eta '}$,
we obtain:
\begin{equation}
S_{\eta '} = \frac{1.2 \pi}{m_{\eta '}} \left(
\frac{2 \pi \Gamma_{\gamma  \gamma} {\cal B} (\eta ' \to \pi^+ \pi^- )}
{m_{\eta '}}  
+ \sin \varphi ' 
\sqrt{2 \pi \sigma_0^{\rm BG} \Gamma_{\eta '} \Gamma_{\gamma  \gamma}
{\cal B} (\eta ' \to \pi^+ \pi^- ) } \right)
\; ,
\label{eqn:bpp}
\end{equation}
where $\sigma_0^{\rm BG}$ is the cross section of the continuum 
$\gamma \gamma \to \pi^+ \pi^-$ component whose amplitude interferes with the
$P$- and $CP$-violating $\eta '$ decay, and
$\varphi '$ is the phase angle and $\sin \varphi ' = -1$
gives the most conservative upper limit of $S_{\eta '}$.

To obtain the upper limit for $S_{\eta '}$ at 90\% confidence level (C.L.),
we have to consider two physically possible cases:
$S_{\eta '}$ is negative or positive, depending on the presence 
or absence of an interference effect between amplitudes of 
opposite $P$ and $CP$.
As the reaction $\gamma \gamma \to \eta '$ would take place via a P wave,
while only even orbital angular momentum waves
can contribute to the ordinary
 $\gamma \gamma \to \pi^+ \pi^-$ process,
it is unlikely that these two processes would interfere.
In that case, $\sigma_0^{\rm BG}=0$ and $S_{\eta '}$ is non negative.
On the other hand, if interference is present,
the lowest boundary of $S_{\eta '}$
is $- \pi \sigma_0^{\rm BG} \Gamma_{\eta '} / 2 \simeq -30$~nb$\cdot$MeV,
where $\sigma_0^{\rm BG} = 93.5$~nb is used, i.e. 
the largest possible value of $\sigma_0^{\rm BG}$ that gives the
most conservative limit.

\begin{figure}
 \centering
 \epsfig{file=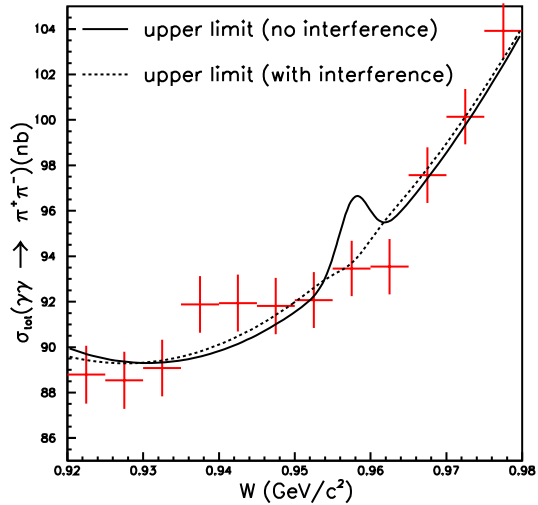,width=70mm}
 \centering
 \caption{Fit result of a Gaussian + second-order polynomial in the 
$\eta ' (958)$ region. The solid (dotted) line is the 90\%~C.L. upper limit
without (with) interference.}
\label{fig:fig10}
\end{figure}

We first obtain $S_{\eta'}^{90}$ , the 90\%~C.L. upper limit 
of $S_{\eta'}$ from the following relation:
\begin{equation}
\int_{S_{\eta'}^{\rm min}}^{S_{\eta'}^{90}} 
\exp \left(- \frac{\chi^2(S_{\eta'})}{2} \right) dS_{\eta'}
 = 0.9 \int_{S_{\eta'}^{\rm min}}^{\infty} 
\exp \left(- \frac{\chi^2(S_{\eta'})}{2} \right) dS_{\eta'} \; ,
\label{eqn:s90}
\end{equation}
where $\chi^2(S_{\eta'})$ is the $\chi^2$ from the fit with a fixed 
$S_{\eta'}$ and
$S_{\eta'}^{\rm min}$ is the lower physical boundary of $S_{\eta'}$.
In the presence (absence) of the interference effect,
the limit is determined
to be $S_{\eta '}^{90} < -2.0~(S_{\eta '}^{90} < 14.4)$~nb$\cdot$MeV.
The results are shown in Fig.~\ref{fig:fig10}.
We obtain the upper limit of ${\cal B}(\eta'\to \pi^+\pi^-)$ taking
the errors of $\Gamma_{\gamma\gamma}$ and ${\cal B}_{\gamma\gamma}$
($\equiv \Gamma_{\gamma\gamma}/\Gamma_{\eta'}$) into account.
Namely, we calculate the contribution to the uncertainty in $S_{\eta '}$ that
arises from these parameters and combine it with the statistical
error of $S_{\eta '}$, $16~{\rm nb}\cdot\MeV$,
reevaluate $S_{\eta '}^{90}$, and then translate it into a limit for
${\cal B} (\eta ' \to \pi^+ \pi^- ) $.
In the case of no interference, we obtain
${\cal B} (\eta ' \to \pi^+ \pi^- ) < 3.3 \times 10^{-4}$.
In the other extreme case of maximum interference, we use
$\sigma_0^{\rm BG} = 93.5$~nb, and
the limit is ${\cal B} (\eta ' \to \pi^+ \pi^- ) < 2.8 \times 10^{-3}$ at
90\%~C.L.  
The errors in $\Gamma_{\gamma\gamma}$ and ${\cal B}_{\gamma\gamma}$ 
are also included but they lead to a negligible change in the upper limits.

\section{Summary and Conclusion}
\label{sec:sum}
In summary, we have performed a high statistics measurement of
the $\gamma \gamma \rightarrow \pi^+ \pi^-$ cross sections in the
$\pi^+ \pi^-$ invariant mass region $0.80~\GeV/c^2 \leq W \leq 
1.5~\GeV/c^2$ for $|\cos \theta^*| < 0.6$ with the Belle detector
at the KEKB $e^+ e^-$ collider.
The total cross section is measured in fine bins of $W$ ($\Diff W=5~\MeV/c^2$) 
and differential cross sections are given in bins of $\Diff W = 5~\MeV/c^2$
and $\Diff |\cos\theta^*| = 0.05$.
We have observed a significant peak corresponding to the $f_0\lr{980}$ 
resonance.
Our data clearly select the peak solution of the Boglione-Pennington
amplitude analysis~\cite{bib:amplitude}.
The total cross section is fitted to obtain the parameters of the
$f_0(980)$ meson~\cite{bib:tmori} and to check consistency in the
$f_2(1270)$ region.
For a $P$- and $CP$-violating decay of the $\eta ' (958)$ meson,
we set an upper limit without (with) interference between
opposite $P$ and $CP$ amplitudes
${\cal B} (\eta ' (958) \to \pi^+ \pi^- ) < 3.3 \times 10^{-4} \;\;
(< 2.9 \times 10^{-3})$
at 90\%~C.L. thereby significantly improving the previous limit of 
0.02~\cite{bib:PDG}.
The angular dependence of the differential cross sections is consistent
with the presence of a significant S wave fraction
for $W \lesssim 1~\GeV/c^2$ and
with the dominance of the D wave in the $f_2(1270)$ mass region.

\section*{Acknowledgment}
We are indebted to M.~Pennington for various enlightening discussions
and useful suggestions.
We thank the KEKB group for the excellent operation of the
accelerator, the KEK cryogenics group for the efficient
operation of the solenoid, and the KEK computer group and
the National Institute of Informatics for valuable computing
and Super-SINET network support. We acknowledge support from
the Ministry of Education, Culture, Sports, Science, and
Technology of Japan and the Japan Society for the Promotion
of Science; the Australian Research Council and the
Australian Department of Education, Science and Training;
the National Science Foundation of China and the Knowledge
Innovation Program of the Chinese Academy of Sciences under
contract No.~10575109 and IHEP-U-503; the Department of
Science and Technology of India; 
the BK21 program of the Ministry of Education of Korea, 
the CHEP SRC program and Basic Research program 
(grant No.~R01-2005-000-10089-0) of the Korea Science and
Engineering Foundation, and the Pure Basic Research Group 
program of the Korea Research Foundation; 
the Polish State Committee for Scientific Research; 
the Ministry of Education and Science of the Russian
Federation and the Russian Federal Agency for Atomic Energy;
the Slovenian Research Agency;  the Swiss
National Science Foundation; the National Science Council
and the Ministry of Education of Taiwan; and the U.S.\
Department of Energy.

\appendix
\section{Background subtraction}
\label{sec:bgds}
In this section, we describe in detail how the background from 
$\eta ' \to \pi^+ \pi^- \gamma$ is subtracted
in the cross section determinations described in Section~\ref{sec:cros} and 
Appendix~\ref{sec:valu}.
Note that beam-gas background that was important in past experiments is 
completely negligible at $B$-factories because of the very high luminosity.
The dominant physics background is due to $\eta '(958)$ production
and its subsequent decay into $\rho^0 \gamma$; the photon
energy at the nominal $\rho^0$ mass is 0.14~GeV, and a significant fraction
of $\pi^+ \pi^-$ pairs satisfy selection requirements such as 
$\mid \sum \vec{p}_t^* \mid < 0.1~\GeV/c$.
The other physics backgrounds are negligibly small.

\begin{figure}[H]
 \subfigure[Background total cross section]
          {\epsfig{file=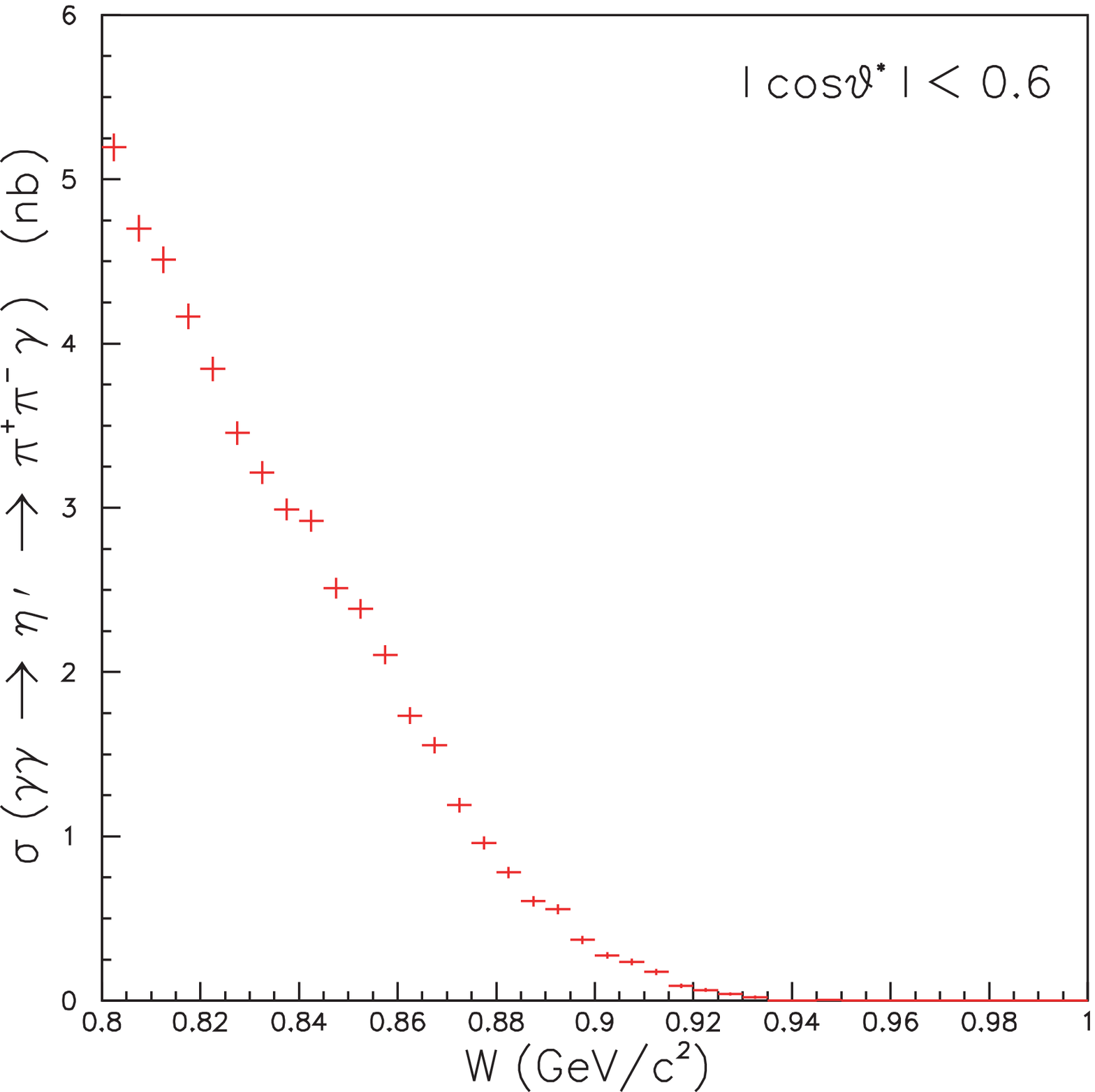,width=70mm}
            \label{fig:fig11a}}
 \subfigure[Background differential cross sections at representative
 $W$'s ($d \sigma (\gamma \gamma \to \eta' \to \rho^0 \gamma \to 
\pi^+ \pi^- \gamma) / d |\cos \theta^*|$ (nb))]
          {\epsfig{file=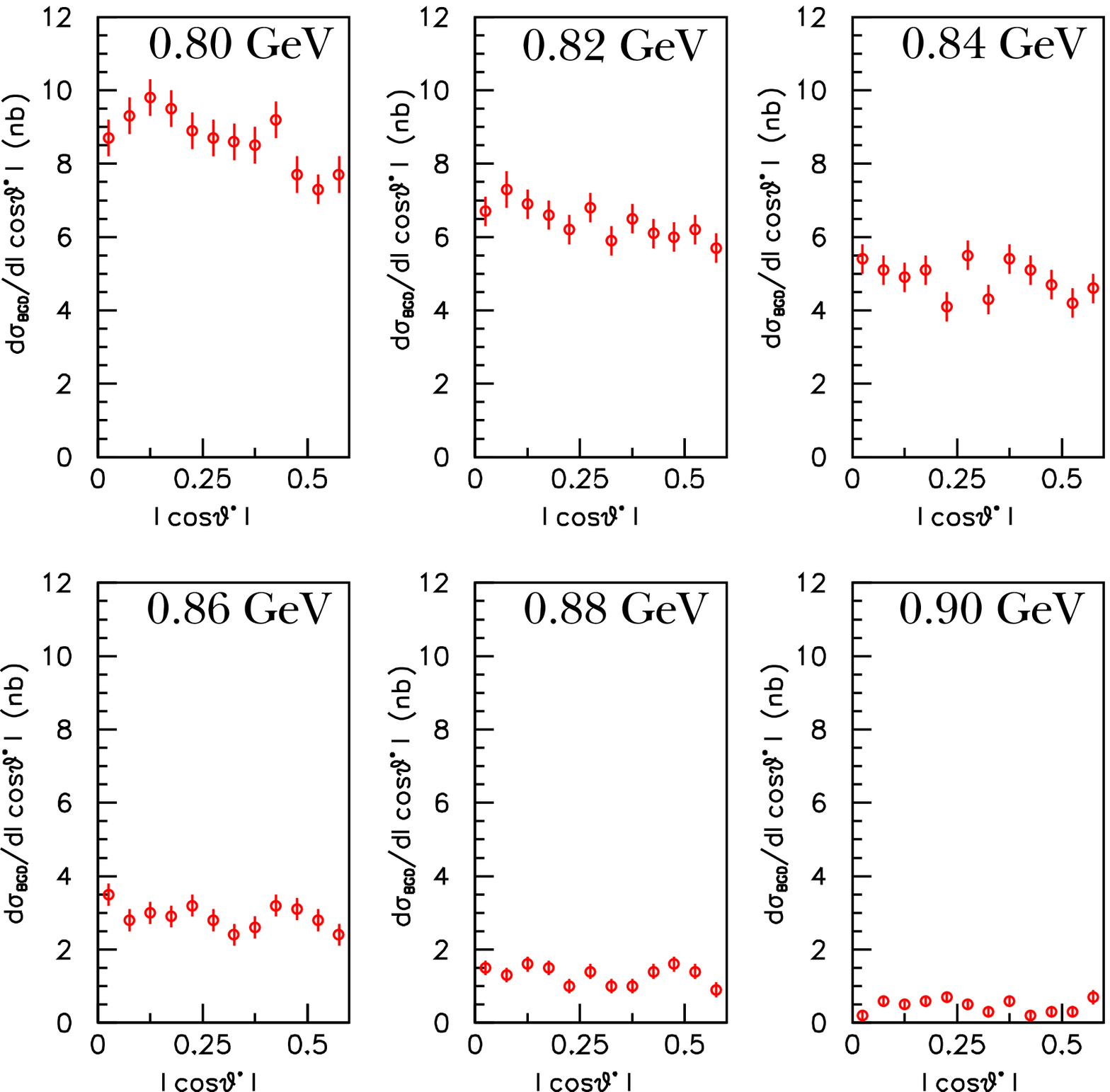,width=70mm}
            \label{fig:fig11b}}
 \caption{Background total and differential cross sections 
($\gamma \gamma \to \eta' \to \rho^0 \gamma \to \pi^+ \pi^- \gamma$)}
\label{fig:fig11}
\end{figure}

The background from $\eta ' \to \pi^+ \pi^- \gamma$ can be estimated with
MC at the four-vector level without doing a full detector simulation.
This is because the subtractions are applied to cross sections where
efficiency corrections of the detector and trigger are already included and
because the bin-sizes used are large enough that
the effect of finite detector resolution is negligible.
The process $e^+ e^- \to e^+ e^- \eta '$ with $\eta ' \to \rho^0 \gamma 
\to \pi^+ \pi^- \gamma$ is simulated using TREPS~\cite{bib:treps}.
The final state $\pi^+ \pi^- \gamma$ is generated
with a matrix element incorporating the dipole transition
feature of the $\eta ' \to \rho^0 \gamma$ decay and the $\rho$ 
pole~\cite{bib:ritten, bib:argus}:
\begin{equation}
|{\cal M} (m_{\pi^+ \pi^-}, E_\gamma , \theta)|^2
\propto \frac{p^2_\pi E^2_\gamma  m^2_{\pi^+ \pi^-} \sin^2 \theta}
{(m^2_\rho - m^2_{\pi^+ \pi^-})^2 + m^2_\rho \Gamma^2(m_{\pi^+ \pi^-})} \; ,
\label{eqn:me}
\end{equation}
where $p_\pi$ is the pion momentum, $E_\gamma$ is the photon energy
and $\theta$ is the angle between one of the pions and the photon, all
evaluated in the di-pion rest system, and the denominator is the
$\rho$ pole.
The mass dependence of the $\rho$ width is parameterized 
as~\cite{bib:jackson}:
\begin{equation}
\Gamma (m) = \frac{2 p^3_\pi}{p_0 (p^2_\pi + p^2_0)} \Gamma_0  \; ,
\label{eqn:rhow}
\end{equation}
where $p_0$ is the $p_\pi$ at $m = m_\rho$ and $\Gamma_0$ is the
nominal $\rho$ width.
Generated $\pi^+ \pi^-$ pairs are subjected to the cut
$\mid \sum \vec{p}_t^* \mid < 0.1~\GeV/c$ and then accumulated into
bins of $W$ - $|\cos \theta^*|$ with the same bin size as that of the 
differential cross sections.
The obtained distribution is related to
the cross section of $\gamma \gamma \to X$ by using~\cite{bib:lum_func}:
\begin{equation}
d \sigma (e^+ e^- \to e^+ e^- X)
= dW \frac{d \cal L}{d W} (W) \sigma (\gamma \gamma \to X)\; .
\label{eqn:eegg}
\end{equation}
The resulting background cross sections to be subtracted
are shown in Figs.~\ref{fig:fig11a} and (b).

\newpage
\section{Total cross section}
\label{sec:valu}
In this appendix, we list the measured values of the total cross section
for the process $\gamma \gamma \rightarrow \pi^+ \pi^-$ integrated over 
the angular region $|\cos \theta^*|<0.6$ in the range of 
$0.8~\GeV/c^2 < W < 1.5~\GeV/c^2$ in steps of $\Diff W = 0.005~\GeV/c^2$.
At each energy, the third number in the table is the statistical
and the fourth is the overall systematic error.

{\small
\begin{table}[h]
\label{tab:total_cross_section}
\begin{supertabular}{|cccc|cccc|cccc|cccc|}
\hline \hline
$W$ & $\sigma$ & stat & syst & $W$ & $\sigma$ & stat & syst &
$W$ & $\sigma$ & stat & syst & $W$ & $\sigma$ & stat & syst \\
 $\GeV/c^2$ & nb & nb & nb & $\GeV/c^2$ & nb & nb & nb &
 $\GeV/c^2$ & nb & nb & nb & $\GeV/c^2$ & nb & nb & nb \\
 \hline
0.8025& 109.74&  2.77&$_{-13.90}^{+14.15}$&
0.8075& 105.40&  1.80&$_{-13.30}^{+13.55}$&
0.8125& 102.85&  1.54&$_{-12.97}^{+13.21}$&
0.8175& 102.25&  1.52&$_{-12.87}^{+13.13}$\\
0.8225& 102.40&  1.48&$_{-12.83}^{+13.08}$&
0.8275&  99.89&  1.44&$_{-12.51}^{+12.77}$&
0.8325& 100.11&  1.43&$_{-12.47}^{+12.71}$&
0.8375&  96.69&  1.41&$_{-12.05}^{+12.30}$\\
0.8425&  94.74&  1.39&$_{-11.82}^{+12.06}$&
0.8475&  97.08&  1.38&$_{-11.96}^{+12.19}$&
0.8525&  93.85&  1.37&$_{-11.58}^{+11.82}$&
0.8575&  90.78&  1.36&$_{-11.22}^{+11.46}$\\
0.8625&  90.40&  1.36&$_{-11.11}^{+11.35}$&
0.8675&  89.21&  1.35&$_{-10.95}^{+11.18}$&
0.8725&  89.02&  1.35&$_{-10.87}^{+11.10}$&
0.8775&  88.96&  1.34&$_{-10.82}^{+11.05}$\\
0.8825&  87.07&  1.33&$_{-10.60}^{+10.83}$&
0.8875&  86.25&  1.32&$_{-10.48}^{+10.71}$&
0.8925&  89.55&  1.32&$_{-10.79}^{+11.02}$&
0.8975&  89.37&  1.31&$_{-10.75}^{+10.97}$\\
0.9025&  85.27&  1.30&$_{-10.32}^{+10.56}$&
0.9075&  89.03&  1.30&$_{-10.69}^{+10.91}$&
0.9125&  87.14&  1.28&$_{-10.49}^{+10.72}$&
0.9175&  90.73&  1.28&$_{-10.84}^{+11.06}$\\
0.9225&  88.69&  1.27&$_{-10.63}^{+10.86}$&
0.9275&  88.47&  1.26&$_{-10.60}^{+10.84}$&
0.9325&  89.00&  1.25&$_{-10.65}^{+10.89}$&
0.9375&  91.84&  1.25&$_{-10.93}^{+11.18}$\\
0.9425&  91.88&  1.25&$_{-10.95}^{+11.20}$&
0.9475&  91.74&  1.24&$_{-10.95}^{+11.20}$&
0.9525&  92.02&  1.23&$_{-10.99}^{+11.26}$&
0.9575&  93.41&  1.22&$_{-11.16}^{+11.42}$\\
0.9625&  93.48&  1.22&$_{-11.19}^{+11.47}$&
0.9675&  97.46&  1.22&$_{-11.62}^{+11.90}$&
0.9725& 100.08&  1.22&$_{-11.89}^{+12.17}$&
0.9775& 103.87&  1.21&$_{-12.29}^{+12.57}$\\
0.9825& 107.07&  1.21&$_{-12.62}^{+12.90}$&
0.9875& 110.24&  1.21&$_{-12.89}^{+13.21}$&
0.9925& 104.68&  1.19&$_{-12.40}^{+12.63}$&
0.9975&  99.79&  1.18&$_{-11.87}^{+12.23}$\\
1.0025& 101.21&  1.17&$_{-12.06}^{+12.36}$&
1.0075& 102.91&  1.17&$_{-12.21}^{+12.51}$&
1.0125& 102.86&  1.16&$_{-12.21}^{+12.51}$&
1.0175& 105.85&  1.16&$_{-12.50}^{+12.80}$\\
1.0225& 104.71&  1.15&$_{-12.39}^{+12.70}$&
1.0275& 107.10&  1.15&$_{-12.64}^{+12.94}$&
1.0325& 108.03&  1.14&$_{-12.75}^{+13.06}$&
1.0375& 109.75&  1.14&$_{-12.94}^{+13.25}$\\
1.0425& 111.16&  1.14&$_{-13.10}^{+13.42}$&
1.0475& 112.96&  1.14&$_{-13.30}^{+13.62}$&
1.0525& 113.58&  1.13&$_{-13.39}^{+13.72}$&
1.0575& 117.81&  1.13&$_{-13.83}^{+14.16}$\\
1.0625& 118.50&  1.13&$_{-13.93}^{+14.27}$&
1.0675& 118.14&  1.12&$_{-13.93}^{+14.29}$&
1.0725& 123.57&  1.13&$_{-14.49}^{+14.85}$&
1.0775& 125.80&  1.12&$_{-14.75}^{+15.12}$\\
1.0825& 127.32&  1.12&$_{-14.95}^{+15.33}$&
1.0875& 132.64&  1.12&$_{-15.51}^{+15.90}$&
1.0925& 135.50&  1.12&$_{-15.85}^{+16.25}$&
1.0975& 137.82&  1.12&$_{-16.14}^{+16.55}$\\
1.1025& 141.86&  1.12&$_{-16.59}^{+17.01}$&
1.1075& 147.31&  1.12&$_{-17.17}^{+17.60}$&
1.1125& 149.05&  1.12&$_{-17.41}^{+17.86}$&
1.1175& 154.11&  1.13&$_{-17.97}^{+18.43}$\\
1.1225& 158.42&  1.13&$_{-18.47}^{+18.94}$&
1.1275& 162.55&  1.13&$_{-18.95}^{+19.45}$&
1.1325& 168.98&  1.14&$_{-19.66}^{+20.16}$&
1.1375& 174.29&  1.14&$_{-20.27}^{+20.79}$\\
1.1425& 179.45&  1.14&$_{-20.87}^{+21.41}$&
1.1475& 185.23&  1.15&$_{-21.53}^{+22.09}$&
1.1525& 190.32&  1.15&$_{-22.13}^{+22.72}$&
1.1575& 196.50&  1.15&$_{-22.83}^{+23.45}$\\
1.1625& 205.00&  1.16&$_{-23.76}^{+24.40}$&
1.1675& 211.52&  1.17&$_{-24.52}^{+25.18}$&
1.1725& 220.50&  1.17&$_{-25.50}^{+26.17}$&
1.1775& 226.11&  1.18&$_{-26.18}^{+26.88}$\\
1.1825& 233.61&  1.18&$_{-27.04}^{+27.76}$&
1.1875& 243.91&  1.19&$_{-28.15}^{+28.89}$&
1.1925& 252.79&  1.20&$_{-29.13}^{+29.89}$&
1.1975& 256.88&  1.20&$_{-29.68}^{+30.46}$\\
1.2025& 265.45&  1.21&$_{-30.62}^{+31.43}$&
1.2075& 273.94&  1.22&$_{-31.55}^{+32.37}$&
1.2125& 280.81&  1.23&$_{-32.32}^{+33.17}$&
1.2175& 286.54&  1.23&$_{-32.97}^{+33.85}$\\
1.2225& 291.98&  1.23&$_{-33.57}^{+34.47}$&
1.2275& 297.12&  1.24&$_{-34.13}^{+35.05}$&
1.2325& 301.35&  1.24&$_{-34.57}^{+35.50}$&
1.2375& 306.15&  1.24&$_{-35.04}^{+35.98}$\\
1.2425& 305.83&  1.24&$_{-35.01}^{+35.97}$&
1.2475& 308.21&  1.24&$_{-35.19}^{+36.14}$&
1.2525& 304.94&  1.24&$_{-34.82}^{+35.78}$&
1.2575& 302.09&  1.23&$_{-34.46}^{+35.40}$\\
1.2625& 297.81&  1.22&$_{-33.93}^{+34.85}$&
1.2675& 290.47&  1.22&$_{-33.08}^{+33.99}$&
1.2725& 281.53&  1.20&$_{-32.05}^{+32.94}$&
1.2775& 271.20&  1.18&$_{-30.87}^{+31.73}$\\
1.2825& 259.63&  1.17&$_{-29.55}^{+30.38}$&
1.2875& 250.20&  1.15&$_{-28.40}^{+29.19}$&
1.2925& 238.08&  1.14&$_{-27.00}^{+27.75}$&
1.2975& 224.97&  1.12&$_{-25.51}^{+26.21}$\\
1.3025& 211.19&  1.10&$_{-23.95}^{+24.62}$&
1.3075& 198.86&  1.09&$_{-22.52}^{+23.15}$&
1.3125& 186.77&  1.07&$_{-21.12}^{+21.71}$&
1.3175& 175.27&  1.06&$_{-19.79}^{+20.34}$\\
1.3225& 164.29&  1.04&$_{-18.52}^{+19.04}$&
1.3275& 152.31&  1.03&$_{-17.18}^{+17.68}$&
1.3325& 141.10&  1.01&$_{-15.94}^{+16.41}$&
1.3375& 132.89&  1.00&$_{-14.98}^{+15.42}$\\
1.3425& 125.24&  0.99&$_{-14.09}^{+14.50}$&
1.3475& 117.89&  0.98&$_{-13.24}^{+13.63}$&
1.3525& 110.82&  0.96&$_{-12.42}^{+12.79}$&
1.3575& 103.72&  0.95&$_{-11.62}^{+11.96}$\\
1.3625&  97.71&  0.94&$_{-10.93}^{+11.27}$&
1.3675&  91.35&  0.93&$_{-10.23}^{+10.48}$&
1.3725&  84.87&  0.92&$_{ -9.50}^{+9.76}$&
1.3775&  81.18&  0.92&$_{ -9.04}^{+9.29}$\\
1.3825&  73.44&  0.91&$_{ -8.23}^{+8.47}$&
1.3875&  70.56&  0.90&$_{ -7.86}^{+8.07}$&
1.3925&  67.25&  0.90&$_{ -7.46}^{+7.67}$&
1.3975&  65.05&  0.90&$_{ -7.17}^{+7.36}$\\
1.4025&  59.39&  0.89&$_{ -6.59}^{+6.77}$&
1.4075&  56.28&  0.88&$_{ -6.23}^{+6.41}$&
1.4125&  52.53&  0.87&$_{ -5.84}^{+6.00}$&
1.4175&  51.69&  0.87&$_{ -5.69}^{+5.85}$\\
1.4225&  48.65&  0.87&$_{ -5.36}^{+5.51}$&
1.4275&  45.78&  0.86&$_{ -5.06}^{+5.20}$&
1.4325&  44.40&  0.86&$_{ -4.88}^{+5.01}$&
1.4375&  41.55&  0.85&$_{ -4.59}^{+4.71}$\\
1.4425&  39.18&  0.84&$_{ -4.34}^{+4.46}$&
1.4475&  38.13&  0.84&$_{ -4.20}^{+4.32}$&
1.4525&  37.14&  0.85&$_{ -4.08}^{+4.18}$&
1.4575&  35.84&  0.84&$_{ -3.93}^{+4.03}$\\
1.4625&  34.07&  0.84&$_{ -3.74}^{+3.84}$&
1.4675&  33.83&  0.84&$_{ -3.68}^{+3.78}$&
1.4725&  31.64&  0.83&$_{ -3.46}^{+3.55}$&
1.4775&  29.58&  0.83&$_{ -3.25}^{+3.34}$\\
1.4825&  29.00&  0.83&$_{ -3.16}^{+3.25}$&
1.4875&  27.52&  0.83&$_{ -3.00}^{+3.08}$&
1.4925&  26.25&  0.82&$_{ -2.85}^{+2.93}$&
1.4975&  25.07&  0.83&$_{ -2.72}^{+2.79}$\\
\hline \hline
\end{supertabular}
\end{table}
}


\begin{thebibliography}{99}
\bibitem{bib:scalar} For a review, see, e.g. C.~Amsler and 
N.A.~T$\ddot{\rm o}$rnqvist, Phys. Rep. 
         \textbf{389}, 61 (2004).
\bibitem{bib:PDG} W.-M.~Yao {\it et al.} (PDG),
                  J. Phys. G {\bf 33}, 1 (2006).
\bibitem{bib:Yang} L.D.~Landau, Sov. Phys. Dokl. \textbf{60}, 207 (1948);
C.N.~Yang, Phys. Rev. \textbf{77}, 242 (1950).
\bibitem{bib:crysball} H.~Marsiske {\it et al.} (Crystal Ball Collaboration), 
Phys. Rev. D \textbf{41}, 3324 (1990).
\bibitem{bib:mark2} J.~Boyer {\it et al.} (Mark II Collaboration), 
Phys. Rev. D \textbf{42}, 1350 (1990).
\bibitem{bib:JADE}T.~Oest {\it et al.} (JADE Collaboration), 
Z. Phys. C - Particles and Fields  \textbf{47}, 343 (1990).
\bibitem{bib:TOPAZ} I.~Adachi {\it et al.} (TOPAZ Collaboration),
Phys. Lett. B \textbf{234}, 185 (1990).
\bibitem{bib:md1}A.E.~Blinov {\it et al.} (MD-1 Collaboration),
Z. Phys. C - Particles and Fields  \textbf{53}, 33 (1992).
\bibitem{bib:CELLO}H. -J. Behrend {\it et al.} (CELLO Collaboration),
Z. Phys. C - Particles and Fields \textbf{56}, 381 (1992).
\bibitem{bib:VENUS} F.~Yabuki {\it et al.} (VENUS Collaboration),
  J. Phys. Soc. Jpn  \textbf{64}, 435 (1995).
\bibitem{bib:amplitude} M. Boglione and M. R. Pennington, Eur. Phys. J.
	 C \textbf{9}, 11 (1999); referred to as BP.
\bibitem{bib:kekb} S.~Kurokawa and E.~Kikutani,
               Nucl. Instrum. and Meth. A {\bf 499}, 1 (2003),
               and other papers included in this volume.
\bibitem{bib:lume} The total cross section for two-photon production of
$\pi^+ \pi^-$ is proportional to 
$\left( \ln (E_{\rm beam}/m_{\pi}) \right)^2 \ln (E_{\rm beam}/m_e)$
in the point-like pion approximation for $E_{\rm beam} \gg m_{\pi}$, where
$E_{\rm beam}$ is the center-of-mass beam energy of an $e^+ e^-$ collider
and $m_{\pi}$ ($m_e$) is the mass of the
pion (electron); see, e.g. S.J.~Brodsky, T.~Kinoshita and H.~Terazawa,
Phys. Rev. Lett.{\bf 25}, 972 (1970).
The first paper to notice the existence of the two photon process appears to
be L.D.~Landau and E.M.~Lifshitz, Sov. Phys. Rev. 
\textbf{6}, 244 (1934).
\bibitem{bib:tmori} T.~Mori {\it et al.} (Belle Collaboration),
Phys. Rev. D {\bf 75}, 051101(R) (2007). Note that the fitted curve
in Fig.~2(a) of this published paper had a slight shift due to an error in 
plotting, which is corrected in Fig.~8(a) in the present paper.  
\bibitem{bib:belle} A.~Abashian {\it et al.} (Belle Collaboration),
                Nucl. Instrum. and Meth. A {\bf 479}, 117 (2002).
\bibitem{bib:geant} R.~Brun {\it et al.}, CERN DD/EE/84-1 (1987).
\bibitem{bib:treps} S.~Uehara, KEK Report 96-11 (1996).
\bibitem{bib:lum_func} V.M.~Budnev {\it et al.}, Phys. Rep.
                 {\bf 15}, 181 (1975).
\bibitem{bib:hel2} see, e.g. H.~Krasemann and J.A.M.~Vermaseren, Nucl. Phys.
B {\bf 184}, 269 (1981).
\bibitem{bib:norm} Note that the normalization of the amplitude is changed
from that of Ref.~\cite{bib:tmori} in order to be consistent with those
for other resonances ($f_2(1270)$ and $\eta '(958)$).
\bibitem{bib:denom} S.M.~Flatt$\grave{\rm e}$, Phys. Lett. {\bf 63B},
                224 (1976); N.N.~Achasov and G.N.~Shestakov, Phys. Rev. 
                 D \textbf{72}, 013006 (2005).
\bibitem{bib:f2shift} A.~Roussarie {\it et al.}, (Mark II Collaboration)
Phys. Lett. B {\bf 105}, 304 (1981).
\bibitem{bib:blat} J.M.~Blatt and V.F.~Weiskopff, 
{\it Theoretical Nuclear Physics} 
(Wiley, New York, 1952), pp.~359-365 and 386-389.
\bibitem{bib:grayer} G.~Grayer {\it et al.}, Nucl. Phys. B {\bf 75}, 
 189 (1974); 
A.~Garmash {\it et al.} (Belle Collaboration), Phys. Rev. D {\bf 71}, 
092003 (2005);
B.~Aubert {\it et al.} (BaBar Collaboration), Phys. Rev. D {\bf 72}, 
052002 (2005).
\bibitem{bib:ritten} A.~Rittenberg, Ph.D. Thesis, LBL, Univ. of California
(1969), ULRL-18863, Sec. VI.A, Table VIII.
\bibitem{bib:argus} H.~Albrecht  {\it et al.} (ARGUS Collaboration), 
Phys. Lett. B {\bf 199}, 457 (1987).
\bibitem{bib:jackson} J.D.~Jackson, Nuovo Cimento {\bf 34},1644 (1964).
\end{thebibliography}
\end{document}